\documentclass[iop, numberedappendix]{emulateapj}

\usepackage{natbib}
\usepackage{multirow}

\shorttitle{The Progenitors of $z\sim2$ Compact, Massive, Quiescent Galaxies}
\shortauthors{Stefanon et al.}

\begin{document}

\title{What Are the Progenitors of Compact, Massive, Quiescent Galaxies at $z=2.3$?\\ The Population of Massive Galaxies at $z>3$ from NMBS and CANDELS}

\author{Mauro Stefanon\altaffilmark{1,2}, Danilo Marchesini\altaffilmark{3}, Gregory~H.~Rudnick\altaffilmark{1}, Gabriel~B.~Brammer\altaffilmark{4} \& Katherine~E.~Whitaker\altaffilmark{5}}

\email{Email: stefanonm@missouri.edu}

\altaffiltext{1}{The University of Kansas, Department of Physics and Astronomy, Malott room 1082, 1251 Wescoe Hall Drive, Lawrence, KS, 66045, USA}
\altaffiltext{2}{Currently at: Physics and Astronomy Department, University of Missouri, Columbia, MO 65211}
\altaffiltext{3}{Physics and Astronomy Department, Tufts University, Robinson Hall, Room 257, Medford, MA, 02155, USA}
\altaffiltext{4}{European Southern Observatory, Alonso de C\'ordova 3107, Casilla 19001, Vitacura, Santiago, Chile}
\altaffiltext{5}{Astrophysics Science Division, Goddard Space Flight Center, Code 665, Greenbelt MD 20771, USA}

\begin{abstract}
Using public data from the NEWFIRM Medium-Band Survey (NMBS) and the Cosmic Assembly Near-Infrared Deep Extragalactic Legacy Survey (CANDELS), we investigate the population
of massive galaxies at $z > 3$. The main aim of this work is to identify
the potential progenitors of $z \sim 2$ compact, massive, quiescent
galaxies, furthering our understanding of the onset and evolution of
massive galaxies. Our work is enabled by high-resolution images from
CANDELS data and accurate photometric redshifts, stellar masses, and
star formation rates (SFRs) from 37-band NMBS photometry. The total number of massive galaxies at $z>3$ is
consistent with the number of massive quiescent galaxies at $z\sim2$,
implying that the SFRs for all of these galaxies must be much lower by
$z\sim2$.  We discover four compact, massive, quiescent galaxies at $z >
3$, pushing back the time for which such galaxies have been observed.
However, the volume density for these galaxies is significantly less
than that of galaxies at $z < 2$ with similar masses, SFRs, and sizes,
implying that additional compact, massive, quiescent galaxies must be
created in the intervening $\sim1$ Gyr between $z = 3$ and $z = 2$.  We find
five star-forming galaxies at $z \sim 3$ that are compact ($R_e < 1.4$ kpc) and have
stellar mass $M_* > 10^{10.6}M_\odot$; these galaxies are likely to become members of the
massive, quiescent, compact galaxy population at $z \sim 2$.  We evolve the
stellar masses and SFRs of each individual $z > 3$ galaxy adopting five
different star formation histories (SFHs) and studying the resulting
population of massive galaxies at $z = 2.3$. We find that declining or
truncated SFHs are necessary to match the observed number density of
massive, quiescent galaxies at $z \sim 2$, whereas a constant SFH would
result in a number density significantly smaller than observed.  All of our assumed SFHs imply number densities of compact, massive, and quiescent galaxies at $z\sim2$ that are consistent with the observed number density. Better agreement with the observed number density of compact, massive, quiescent galaxies at $z\sim2$ is obtained if merging is included in the analysis and better still if star formation quenching is assumed to shortly follow the merging event, as implied by recent models of formation of
massive, quiescent galaxies.
\end{abstract}


\section{Introduction}

The population of galaxies in the Local Universe presents a clear bi-modality, as evidenced by color-magnitude diagrams (e.g \citealt{baldry2004,kauffmann2004}), with galaxies either living on the red sequence or in the blue cloud.  This bi-modality is further supported by tight correlations between the main physical properties of each class (e.g. \citealt{tully1977,kormendy1989}).  Galaxies on the red sequence typically are massive and quiescent (i.e. with low or no ongoing star formation) with early-type morphologies, while galaxies in the blue cloud are less massive, with higher star formation rates (SFRs) and have spiral or irregular morphologies. 

The physical mechanisms driving the onset of the observed bi-modality are one of the main open issues in the study of galaxy formation. At high redshift the observational picture is complicated by the lack of a common definition for massive, quiescent galaxies (MQGs) in the literature (see discussion in \citealt{saracco2012}). 

Samples selected according to the spectral energy distribution (SED) show that massive (i.e. stellar mass $M_*\gtrsim 10^{11}M_\odot$) and quiescent (i.e. specific star formation rate sSFR $\lesssim 10^{-11}-10^{-10}$yr$^{-1}$) galaxies were already in place at $z\sim2$ \citep{franx2003,daddi2005, kriek2006,cimatti2008}. Their number density has grown by almost a factor of 10 since $z=3$ \citep{labbe2005,arnouts2007,fontana2009,taylor2009,ilbert2010,cassata2011,brammer2011,dominguez-sanchez2011,bell2012}, although most of the evolution occurred at $z > 1$ (e.g. \citealt{pozzetti2007}).  

In addition to rapid number density evolution, the sizes of MQGs have evolved dramatically from high redshift to the present day. Most of MQGs at $z>1.5$ are also compact systems: the observed effective radii are generally a factor of $\sim3-5$ smaller at $z\sim2.5$ \citep{daddi2005,longhetti2007,toft2007,trujillo2007,vandokkum2008,cimatti2008,saracco2009,vanderwel2008,bezanson2009,szomoru2012} than at $z=0$, where such compact galaxies are almost entirely absent \citep{trujillo2009,taylor2010}.

There have been multiple attempts to explain the evolution of these objects and to place them into a broader cosmological context. In the so-called monolithic collapse scenario, MQGs would have assembled almost all of their stellar mass at high redshift, followed by a passive evolution of the stellar population. This class of models, however, foresees little or no evolution in size, in contrast to observations. More recent models describe the formation of MQGs as a two-stage process: gas-rich merger events generate compact, massive spheroids at $z \gtrsim 3$, while minor, dry (i.e. without formation of new stars) mergers at later cosmic times  would increase the size, while responsible for only a small increase in stellar mass \citep{naab2007,naab2009,wuyts2010,oser2012}. Indeed, \citet{bezanson2009} showed that the stellar mass density of the inner $\sim1$kpc of $z\sim2$ galaxies does not substantially differ from that in present-day MQGs, suggesting that the growth during the second phase progresses from the inner region rapidly towards the external parts (the so-called inside-out growth - see e.g. \citealt{vandokkum2010}).

However, the picture is still far from being completely clear: detailed comparisons with $\Lambda$CDM models of dry merging show that some of the models predict descendants of $z>2$ too compact compared to the observed local MQGs \citep{cimatti2012}. Recent studies have also revealed the existence of a large number of MQGs at $z\sim1.5$ with sizes similar to those of the local galaxies with comparable mass \citep{saracco2009,saracco2010,mancini2010}.  If the population of MQGs at $z\gtrsim2.5$ includes only compact systems, this would imply a size evolution timescale of $\approx 1$ Gyr, challenging the current models of formation of local MQGs \citep{saracco2010}.

The main goal of this work is to identify, within an observational framework, the population of galaxies at $z>3$ which could give rise to the population of compact, MQGs observed at $z=2.3$. We combine data from two public surveys, the NEWFIRM Medium-Band Survey (NMBS, \citealt{whitaker2011}) and the Cosmic Assembly Near-infrared Deep Extragalactic Legacy Survey (CANDELS, \citealt{grogin2011, koekemoer2011}). The NMBS provides accurate measurements of photometric redshifts ($z_{\rm phot}$), stellar masses ($M_*$) and star-formation rates (SFR), while robust morphologies and sizes were measured from the high-resolution imaging from CANDELS.

In Section \ref{sec:data}, we describe the data sets used in our work, the measurements of the sizes, and the completeness in stellar mass.  The results are presented in Section \ref{sec:results}. We discuss our results in Section \ref{sec:learn}, and summarize them in Section 5.

Throughout this work, we use a concordance cosmology with $H_0=70$ Km s$^{-1}$ Mpc$^{-1}$, $\Omega_m=0.3$ and $\Omega_\Lambda=0.7$. All magnitudes are referred to the AB system.

\section{Data}
\label{sec:data}

We combined the photometric depth and the accurate measurements of photometric redshifts offered by the NMBS  (\citealt{vandokkum2009,whitaker2011}) with the high-resolution HST/WFC3 deep NIR imaging from the CANDELS (\citealt{grogin2011,koekemoer2011}). These two surveys will be briefly described in the next two sections.

\subsection{NMBS-COSMOS}

The NMBS covers two fields: a $27.6" \times 27.6$"  pointing within the COSMOS field \citep{scoville2007} and a second pointing, of the same size, overlapping with part of the All-Wavelength Extended Groth Strip International Survey (AEGIS) strip \citep{davis2007}. Only the data from the COSMOS field was used, since the overlap between NMBS-AEGIS and CANDELS is minimal; using both fields the average improvement in cosmic variance would have been just 5\%.

NMBS uses five medium-bandwidth filters in the NIR wavelength range $1-1.8{\rm~}\mu$m. The bluest filter is similar to the Y band, the canonical J  and H bands are split into two filters each. Such configuration pinpoints the location of the redshifted Balmer/$4000$~\AA breaks in $1.5<z<3.5$ galaxies \citep{vandokkum2009} and enables photometric redshift measurements with an accuracy of  $\sigma_z /(1 + z) \sim 2\%$ for objects in the redshift range $1.5<z<3.5$ \citep{whitaker2011}.

The full public catalog for the COSMOS field \citep{whitaker2011}\footnote{Catalogs can be downloaded from http://www.astro.yale.edu/nmbs/Data\_Products.html} provides UV-to-24~$\mu$m photometry for 31306 K-selected sources based on the de-blended version, along with accurate photometric redshifts, rest-frame luminosities, star formation rates, and stellar mass measurements. The redshifts and stellar masses were computed using 37 filters from the COSMOS fields, combining the NIR medium-bandwidth data with existing UV (Galaxy Evolution Explorer), visible and NIR (Canada-France-Hawaii Telescope and Subaru Telescope), and mid-IR (Spitzer/IRAC) imaging \citep{whitaker2011}. 

\begin{figure*}
\epsscale{1.0}
\includegraphics[bb = 80 640 700 800,clip]{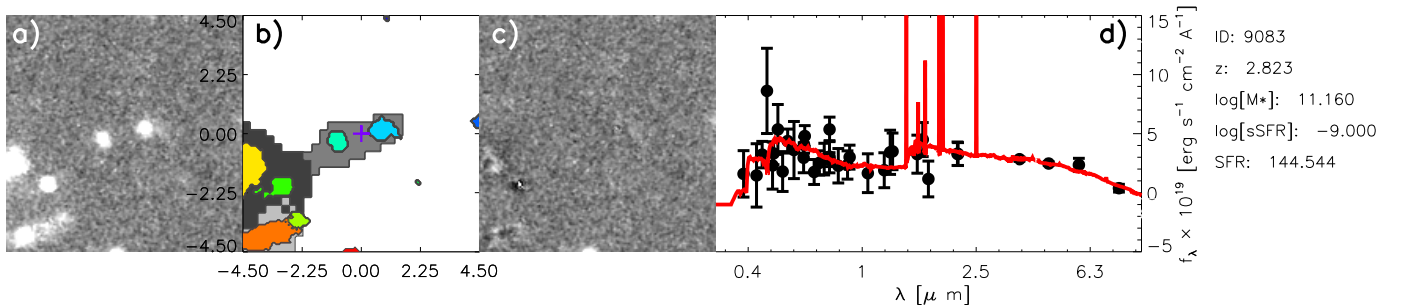}
\caption{Example of the need for high-resolution images such as those from CANDELS in identifying blended objects in lower resolution images. From left to right: a) Tile from the F160W CANDELS image; b) Segmentation map from NMBS (grey scale) and from the F160W CANDELS frame (colored spots); the purple cross indicates the center of the NMBS source, which is the center of the frame; the axis values are in arcseconds relative to the center of the tile; c) Residual image from GALFIT after fitting all sources simultaneously; the pixel scale for tiles a), b) and c) is the same. d) Observed SED from NMBS (black points with error bars) and SED fit from EAZY (solid red line). The blended objects like the one here plotted were excluded from the analysis. \label{fig:deblending}}
\end{figure*}

The catalog is complete at the 75\% detection level for magnitudes brighter than $K_S=23.1$AB \citep{whitaker2011}; we selected objects brighter than $K_S=23.4$ mag, corresponding to a detection completeness of 50\%. 

\subsection{CANDELS}

CANDELS  is a 902-orbit Hubble Space Telescope (HST) Multi-Cycle Treasury program aimed at probing the evolution of galaxies and black holes from $z\sim 1.5$ to 8 and at detecting and studying Type Ia supernovae at $z > 1.5$ in order to better constrain the nature of dark energy \citep{koekemoer2011}.

The Wide portion of the survey covers a total of  $\sim800$ square arcminutes down to $H\simeq 26.5$ mag, spread over five fields: extended regions around the two GOODS fields,  the Extended Groth Strip (EGS: \citealt{davis2007}), COSMOS \citep{scoville2007}, and the UKIDSS Ultra-Deep Survey (UDS: \citealt{lawrence2007}).

We used the F160W as the filter for the measurement of the morphological parameters as it is the reddest band available with high-resolution imaging. At redshift $z\sim3$, the F160W filter corresponds to the rest-frame B-band.\\

\subsection{Sample selection}

The Full Width Half Maximum (FWHM) of the Point-Spread Function (PSF) on the CANDELS F160W images  is FWHM$\simeq 0.17"-0.19"$  \citep{koekemoer2011}, a factor of $\sim$6 better than the $K$-band PSF in NMBS. This fact, together with the higher photometric depth compared to the NMBS survey, translates into a higher surface brightness sensitivity than NMBS.

From the full NMBS catalog we extracted those non-stellar objects and objects with a good quality flag (\emph{star\_flag=0} and \emph{use=1}). We selected galaxies compatible with being at $z>3$ at the 1-sigma level (i.e. \emph{u\_68} $> 3.$). In this way, we selected a total of 613 galaxies. 

\begin{figure}
\epsscale{1.0}
\plotone{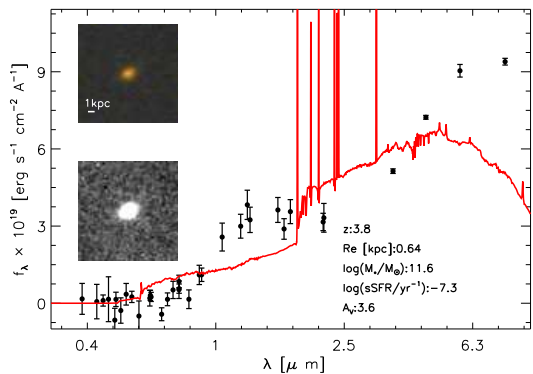}
\caption{Observed SED from NMBS (black points), EAZY best-fitting SED (solid red curve) for a possible AGN. A color cutout created from the F814W, F125W and F160W CANDELS filters is shown in the top-left corner, while a cutout from CANDELS F160W image is presented at the bottom left. The angular size for both cutouts is $3.7"\times3.7"$.Objects like the one plotted were excluded from the sample. \label{fig:agn}}
\end{figure}

The overlap of the CANDELS COSMOS image and the NMBS COSMOS field amounts to 192 square arcmin, $\sim 20\%$ of the original NMBS field. This reduces the sample of $z>3$ galaxies available to the measurement of the morphological parameters to 133 sources.

The availability of CANDELS data allowed us to accurately measure the morphological parameters of our $z>3$ galaxies, and to identify potentially blended objects, which appear as single sources in the NMBS catalog. In Figure \ref{fig:deblending} we show an example of the importance of high-resolution imaging, which allowed us to identify those cases in which blended and/or very close objects could affect the measurement of the SED, with consequently unreliable photometric redshifts and stellar population properties (e.g. stellar masses, SFRs). For these reasons, blended objects were excluded from our final analysis. We also visually inspected the 133 SEDs of the $z>3$ galaxies and excluded from the sample those showing the presence of a possible AGN ( an example of which is presented in Figure  \ref{fig:agn}). The fraction of such objects in the whole sample of 133 galaxies summed to 4\%.

\begin{figure}
\epsscale{1.0}
\hspace{-0.5cm}
\includegraphics[width=9.2cm]{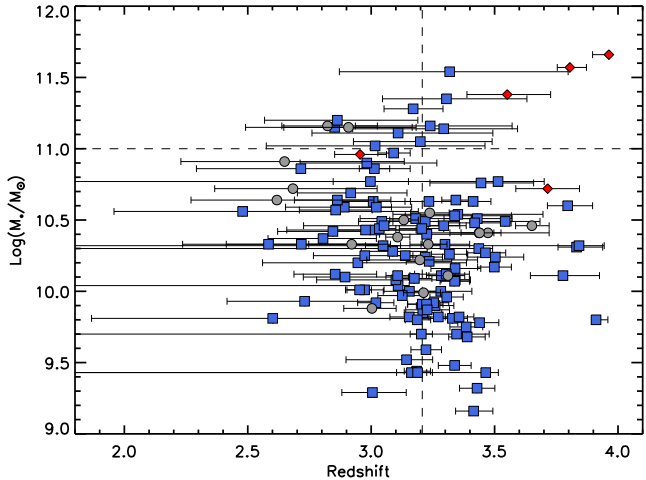}
\caption{Stellar mass as a function of redshift for the 110 galaxies constituting our final sample (blue filled squares), after cleaning for blended objects (grey filled circles) and possible AGN (red filled diamonds). The error bars in redshift encompass the 68\% confidence level, as estimated by EAZY. The median redshift of $M_*>10^{11}M_\odot$ galaxies is 3.2, and it is marked by the vertical dashed line. All $M_*>10^{11} M_\odot$ galaxies, selected to possibly lie at $z>3$, have a redshift $z>2.5$ at a 68\% confidence level. This gives us confidence that our selection is not including $z\sim2$ galaxies with very broad redshift probability distributions.\label{fig:m_star-z}}
\end{figure}

The final catalog includes 110 galaxies. The distribution of $M_*$ with $z_{\rm phot}$ for the final sample is presented in Figure \ref{fig:m_star-z}. The median redshift and redshift uncertainty for the full sample are $z_{\rm tot}=3.2$ and 0.1 respectively; similarly, for the $M_*>10^{11}M_\odot$ subsample, we have $z_{\rm M}=3.2\pm0.3$. The plot shows also that out of the 10 objects with redshift compatible with  $z\sim2.3$, none of them has a stellar mass  $M_*>10^{11}M_\odot$, increasing our confidence in the adopted redshift selection criteria.

\subsection{Size measurement}

\label{sec:size}

The luminosity profile of each galaxy was fit by a single \citet{sersic1968} profile, using the GALFIT \citep{peng2002,peng2010} program. This code fits two-dimensional analytic functions directly to images after convolving the profile with a PSF. All size measurements were done on the F160W CANDELS image. 

\begin{figure}
\epsscale{1.0}
\hspace{-0.5cm}
\includegraphics[width=9.2cm]{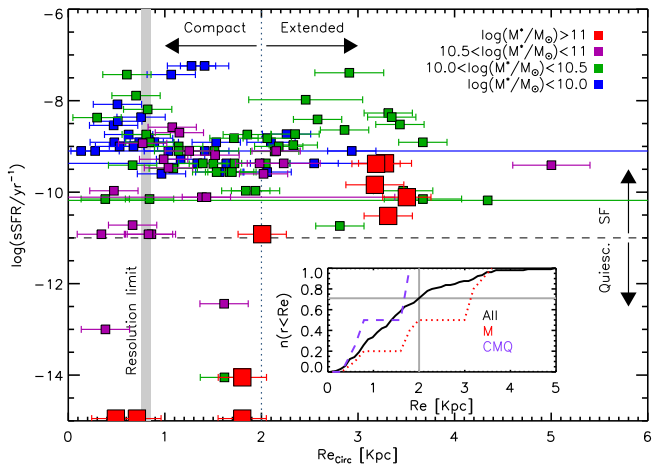}
\caption{Specific star formation rate as a function of $R_{e}$. Associated errors on $R_e$ are from GALFIT. Points are color-coded according to their stellar mass. The sample defined by $\log(M_*/M_\odot)>11$ and marked in the plot by the large red filled squares, is complete in stellar mass at the 70\% level. As such, this is the sample we adopt to draw our conclusions. The vertical grey region marks the limit to which we can resolve objects, determined by fitting stars with a \citet{sersic1968} profile.  We note that, given the existence of a size-mass relation, the vertical line at $R_e=2$~kpc marking the separation between compact and extended galaxies only applies to $\log(M_*/M_\odot)>11$ galaxies (see Section \ref{sect:cmsf} for a more complete discussion). The inset shows the cumulative fraction of galaxies according to their effective radius $R_e$ for the full sample (solid black line), for the massive ($\log(M_*/M_\odot)>11$) galaxies (dotted red line) and for the compact ($R_e<2$~kpc), massive, quiescent ($\log({\rm sSFR/yr}^{-1})<-11$) sample (dashed purple line). A vertical grey line marks our limit in $R_e$ for the compact galaxies. Approximately 70\% of the galaxies in our (flux-limited) sample have $R_e<2$~kpc; 4 out of the 10 massive galaxies are also compact. No extended quiescent galaxies are present, although this could be the result of a surface brightness effect. Star-forming galaxies show no evident correlation between $R_e$ and sSFR. \label{fig:re-ssfr}}
\end{figure}

A zero-order set of morphological parameters was obtained via SExtractor \citep{bertin1996} on the F160W image. In order to make the measurements of the morphological parameters with GALFIT as robust and reliable as possible, two fundamental steps were taken into account. The first is the construction of the PSF, while the second is the evaluation of the sky background level. 

The PSF was constructed with IRAF/Daophot \citep{stetson1987}, from a set of unsaturated, bright and isolated stars, and using a gaussian analytic function plus a look-up table built by the IRAF \texttt{psf} task from the residuals of the function fitting. This allowed us to take into account the anisotropies of the brightness profile, especially those in the wings of the PSF.

The SExtractor sky measurement is based on thresholding: the sky is measured by determining when the gradient of the flux flattens out enough compared to the background noise. In particular this means that SExtractor will generally over-predict the sky with the effect of suppressing the S\'ersic index,  the effective radius, and luminosity. In our analysis, the background level was estimated from its  median value in 30  non-overlapping boxes 50 pixels (3") wide, distributed across a region of 400 pixels (24") around the central object and free from any other source, as probed by the SExtractor segmentation map. The region size is wide enough to grant robust statistics and at the same time it is small enough to mitigate possible gradients in the sky background. The boxes were kept to an additional distance of 0.6" from the segmentation map, increasing the confidence in excluding contamination from the outskirts of objects.

All objects in a box of  150 pixels (9") around each galaxy were simultaneously fit, in order to take into account possible contamination of the brightness profile from neighboring objects, which could bias the measured size and S\'ersic index.

The effective radii were finally circularized and converted to physical units using the adopted cosmology.

The resulting distribution of $R_e$  as a function of the sSFR for the galaxies of our sample are presented in Figure \ref{fig:re-ssfr}.

\subsection{Stellar Mass Completeness}

Consistent statistical measurements of intrinsic physical quantities rely on the accurate characterization of the selection effects, i.e. on the completeness of the sample.

The galaxy stellar mass completeness for a flux-limited sample not only directly depends on the limiting flux itself, but also on the mass-to-light ratio ($M/L$) of each galaxy. All else being the same, galaxies with lower $M/L$ values will be probed to lower mass limits; similarly, it is possible to probe only the higher stellar mass region for those object with higher $M/L$ ratios (see e.g. \citealt{marchesini2009}).

The above effect makes the measurement of the stellar-mass completeness a non trivial task; common approaches involve either SSP modelling \citep{dickinson2003} and/or comparing the galaxy populations to existing deeper data \citep{marchesini2010}.

\begin{figure}
\epsscale{1.0}
\hspace{-0.5cm}
\includegraphics[width=9.2cm]{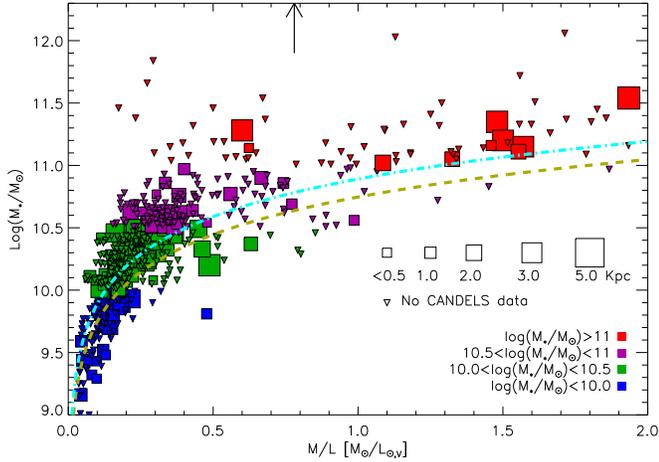}
\caption{Stellar mass as a function of the rest-frame V-band mass-to-light ratio for all the NMBS galaxies with redshift compatible with being at $z>3$. Points are color-coded according to their stellar mass. Objects covered by  the CANDELS F160W frame which allowed for size measurement are marked by larger symbols, with the symbol size proportional to the $R_e$. The 50\% completeness level, computed adopting the $K$-band detection completeness curve from \citet{whitaker2011}, is marked by the yellow dashed curve, while the 80\% completeness level by the cyan dash-dotted curve. The vertical arrow marks the $M/L$ of a single stellar population passively evolved to $z=3.2$, smaller than the maximum value we observe for the massive population, increasing our confidence on our completeness analysis. The completeness of the $M_*>10^{11} M_\odot$ galaxies  is $>70\%$. \label{fig:compl}}
\end{figure}

The average PSF on NMBS frames is $\sim1.1"$, corresponding to an $R_e$ of $\sim 0.6"$, or $\sim$5~kpc at $z=3.2$. In particular this means that objects with $R_e<5$~kpc are essentially point sources on the NMBS frames. The inset in Figure~\ref{fig:re-ssfr} shows that the totality of the galaxies in our sample has $R_e<5$~kpc. We therefore adopt, for each galaxy, the $K-$band completeness curve measured for point sources for the NMBS survey and presented in \citet{whitaker2011}. 

The completeness of the $z>3$ sample  as a function of stellar mass was analyzed for different ranges in $M/L$ ratio. Objects with $\log(M_*/M_\odot)>11$ are complete at the $\sim 70\%$ level. A graphical representation is plotted in Figure \ref{fig:compl}, which shows the $\log (M_*/M_\odot)$ vs.  $M/L$ ratio diagram for all NMBS galaxies at $z>3$, as well as the 50\% and 80\% completeness curves of the stellar mass as a function of $M/L$ ratio derived from the $K$-band  detection completeness curve from \citet{whitaker2011}. Figure \ref{fig:compl} shows that for galaxies with $\log (M_*/M_\odot)>11$, the completeness is $\gtrsim 70\%$. Also plotted is the $M/L$ at $z=3.2$ of a passively evolving single stellar population formed at $z_{\rm form}=20$. Its value, equal to $0.78 M_\odot/L_{\odot,V}$, well below the maximum value we observe for the massive sample, provides additional confidence in the adopted completeness limits in stellar mass. We note furthermore that the completeness in stellar mass as a function of redshift is fairly flat over the redshift range targeted in our work, changing by, e.g., $\sim0.1$~dex from $z=4$ to $z=3$ for a single stellar population formed at $z_{\rm form}=10$. In our analysis we then selected only those galaxies with $\log(M_*/M_{\odot})>11$ (for a total of 10 objects), for which small completeness corrections are required; specifically, the correction for incompleteness ranged from 1.05 to 1.5.

\begin{figure*}
\epsscale{1.0}
\begin{tabular}{ccc}
\includegraphics[width=5.6cm]{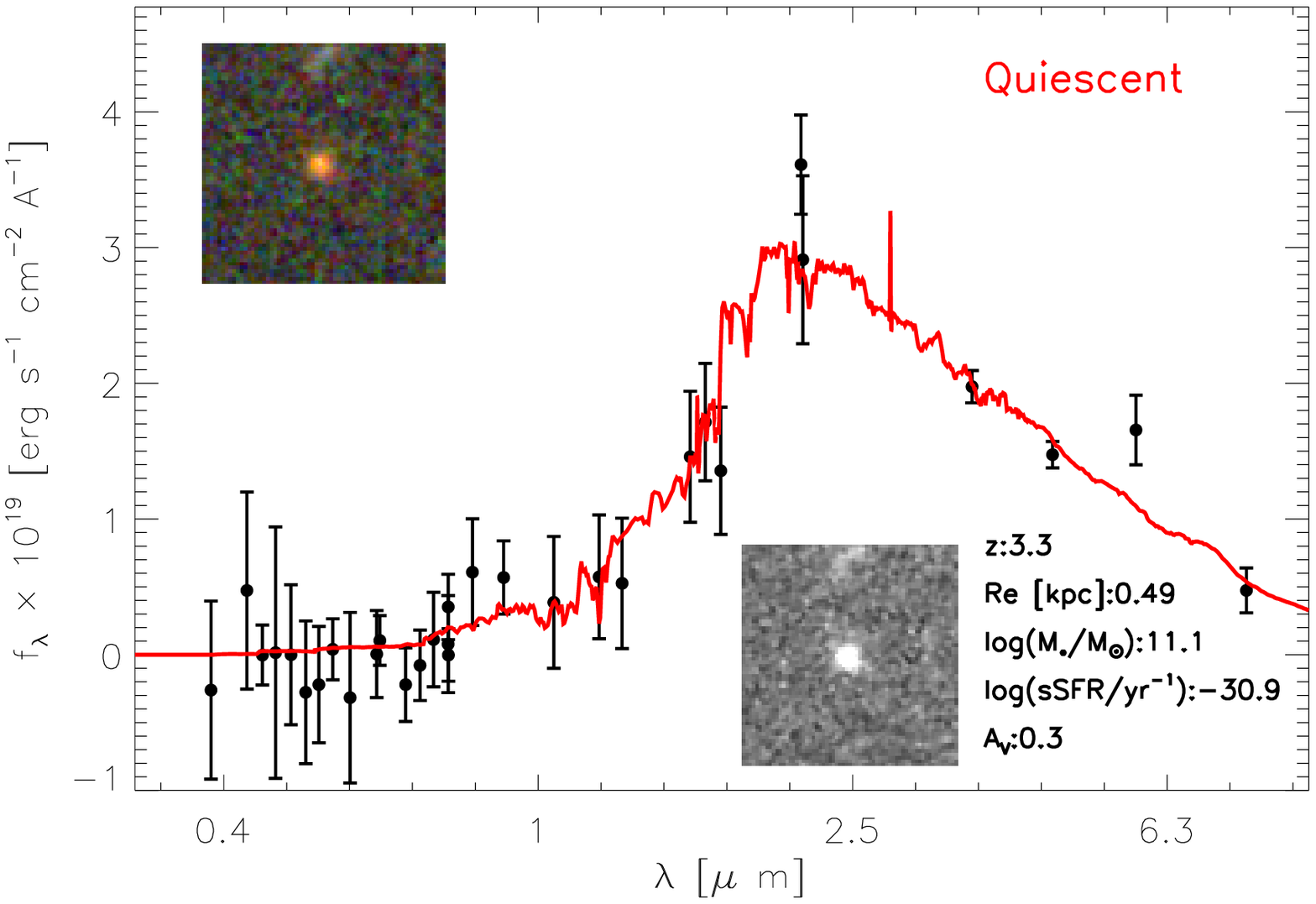} &  \includegraphics[width=5.6cm]{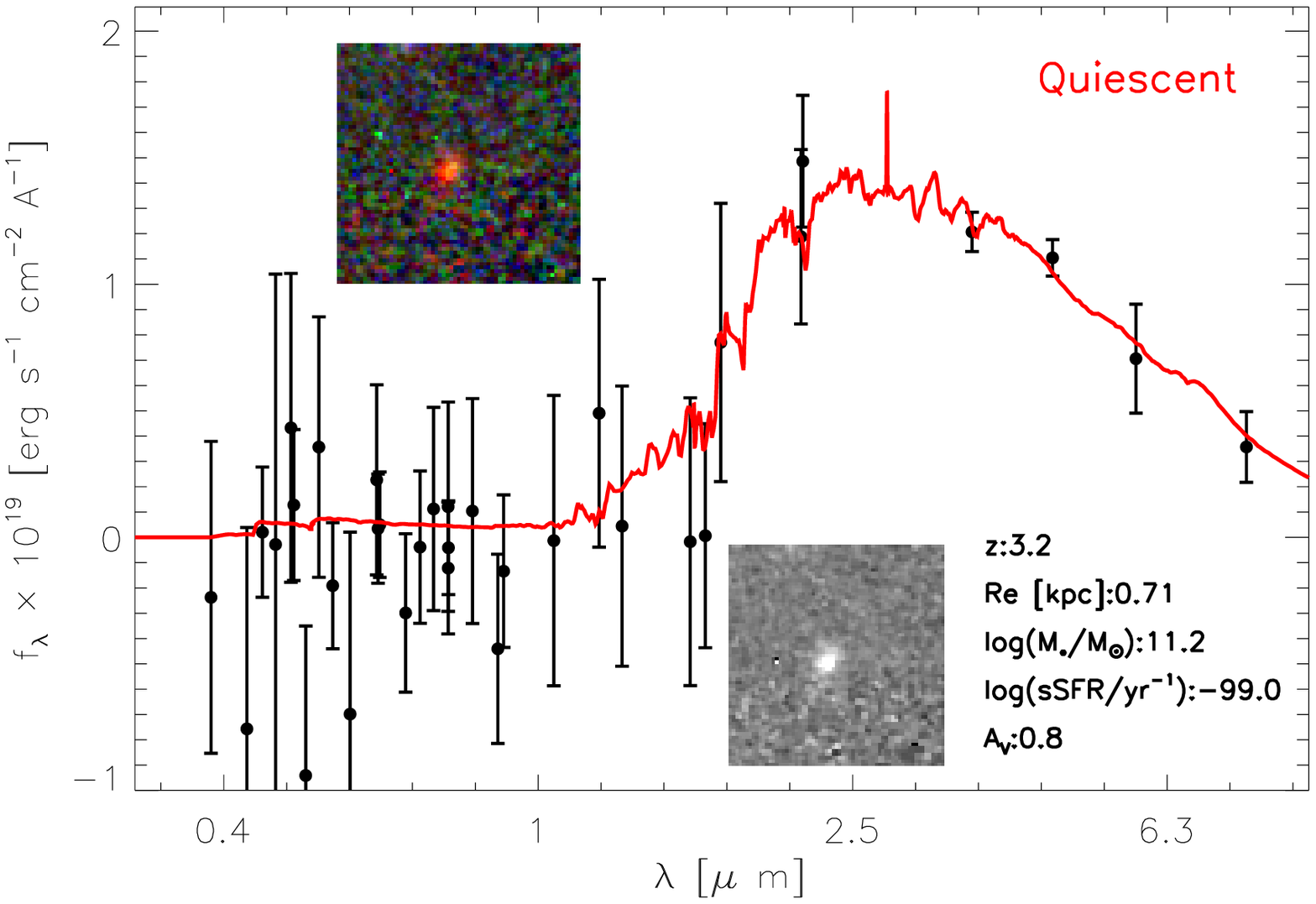} &  \includegraphics[width=5.6cm]{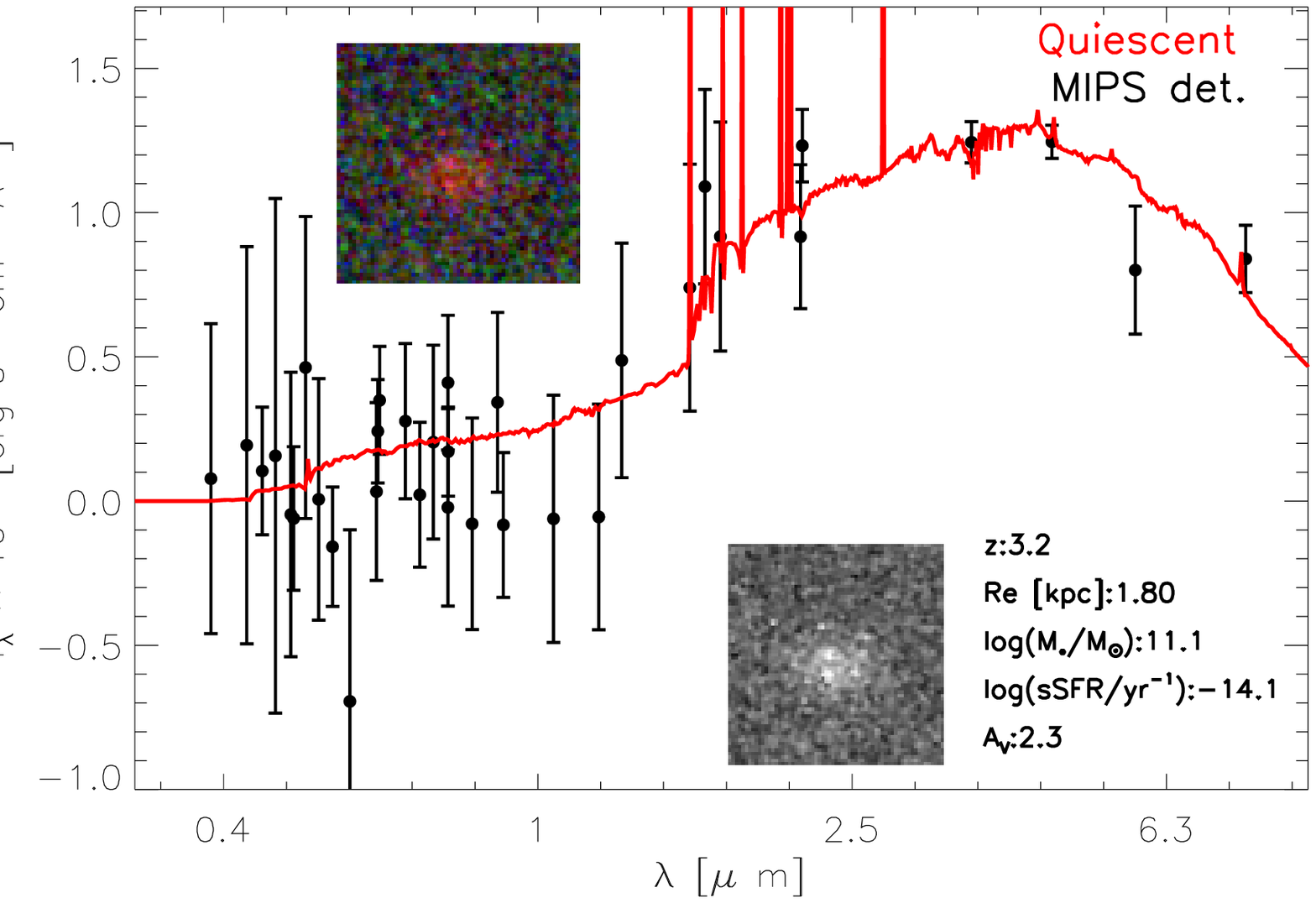} \\
\includegraphics[width=5.6cm]{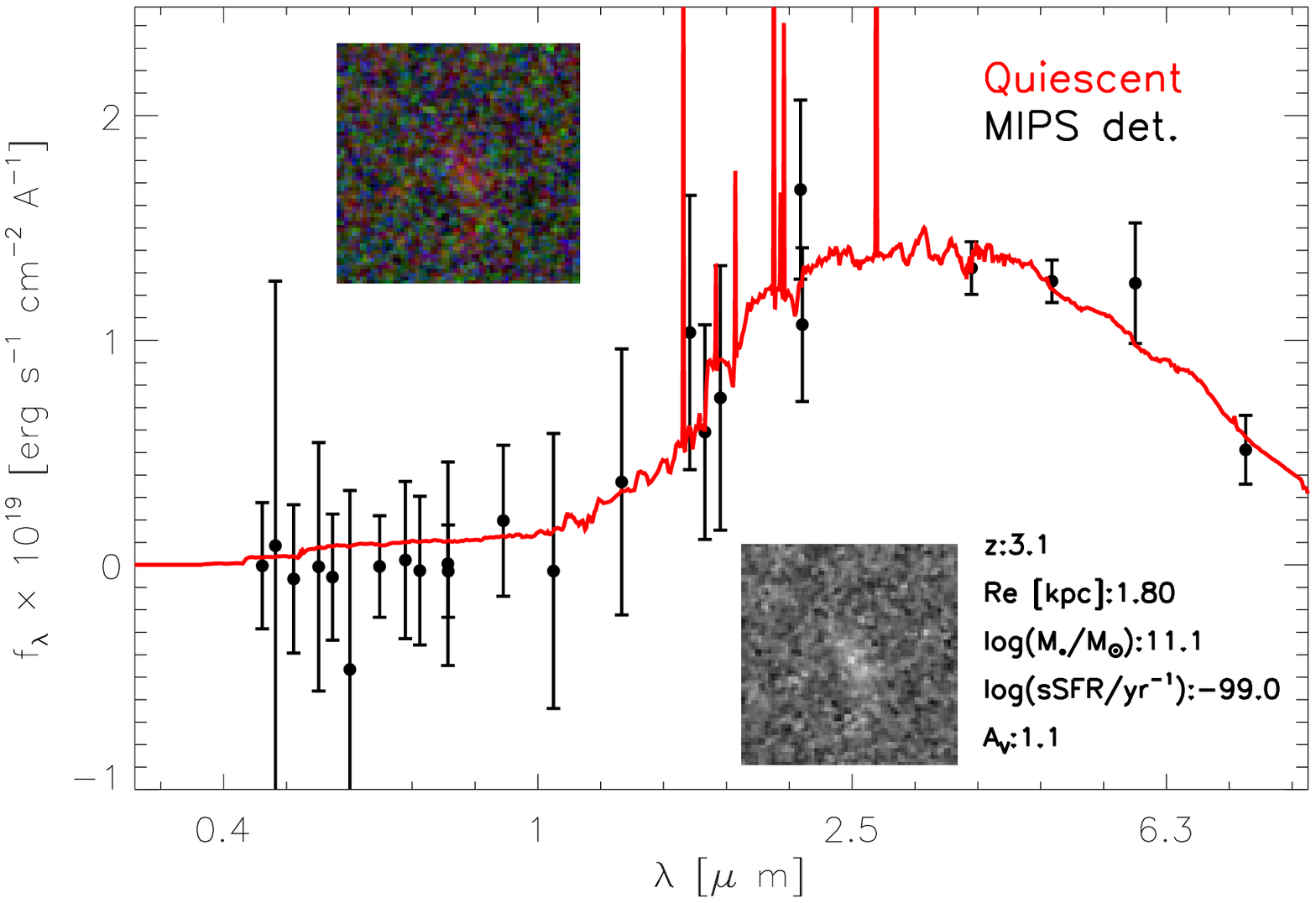} &  \includegraphics[width=5.6cm]{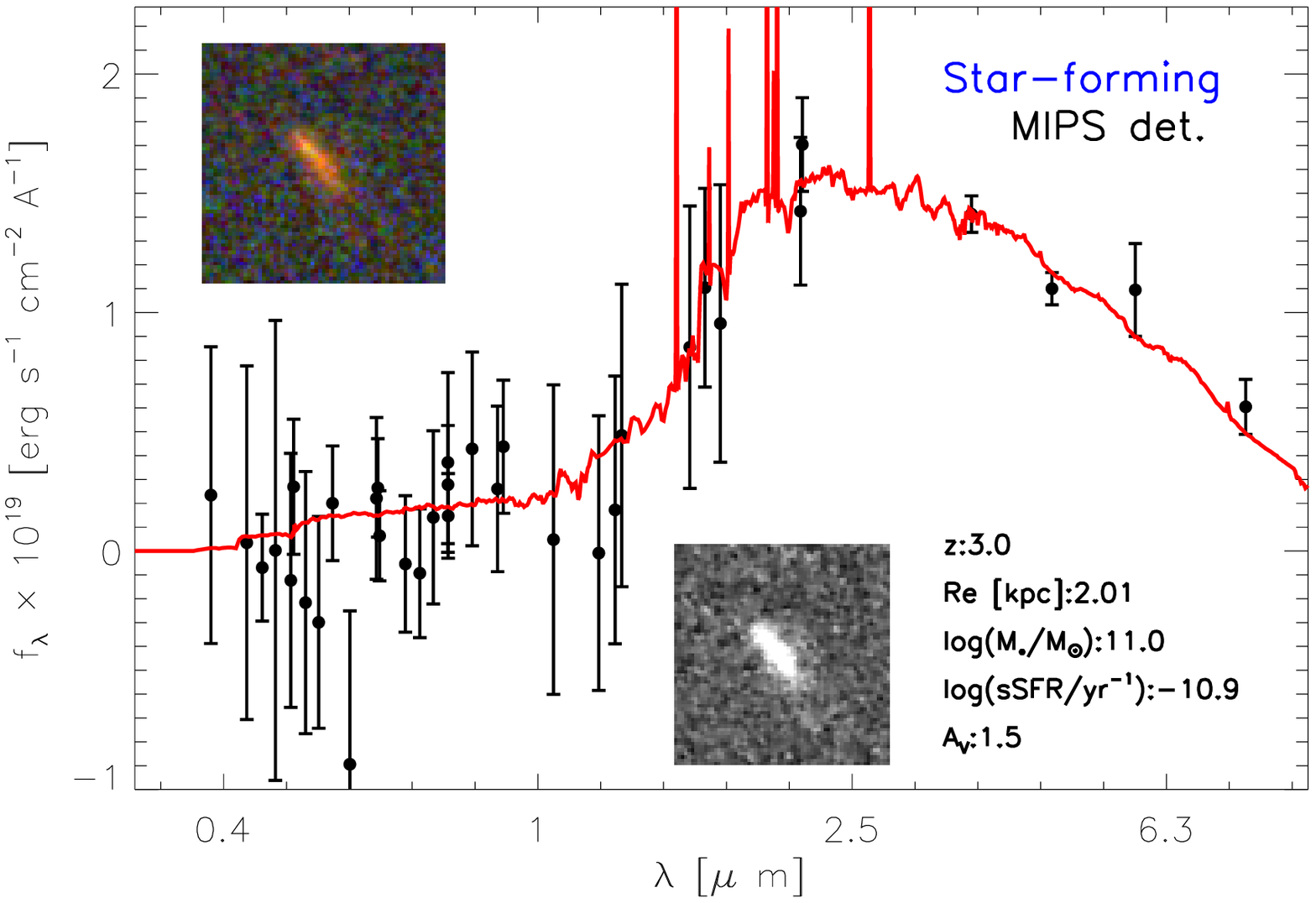} &  \includegraphics[width=5.6cm]{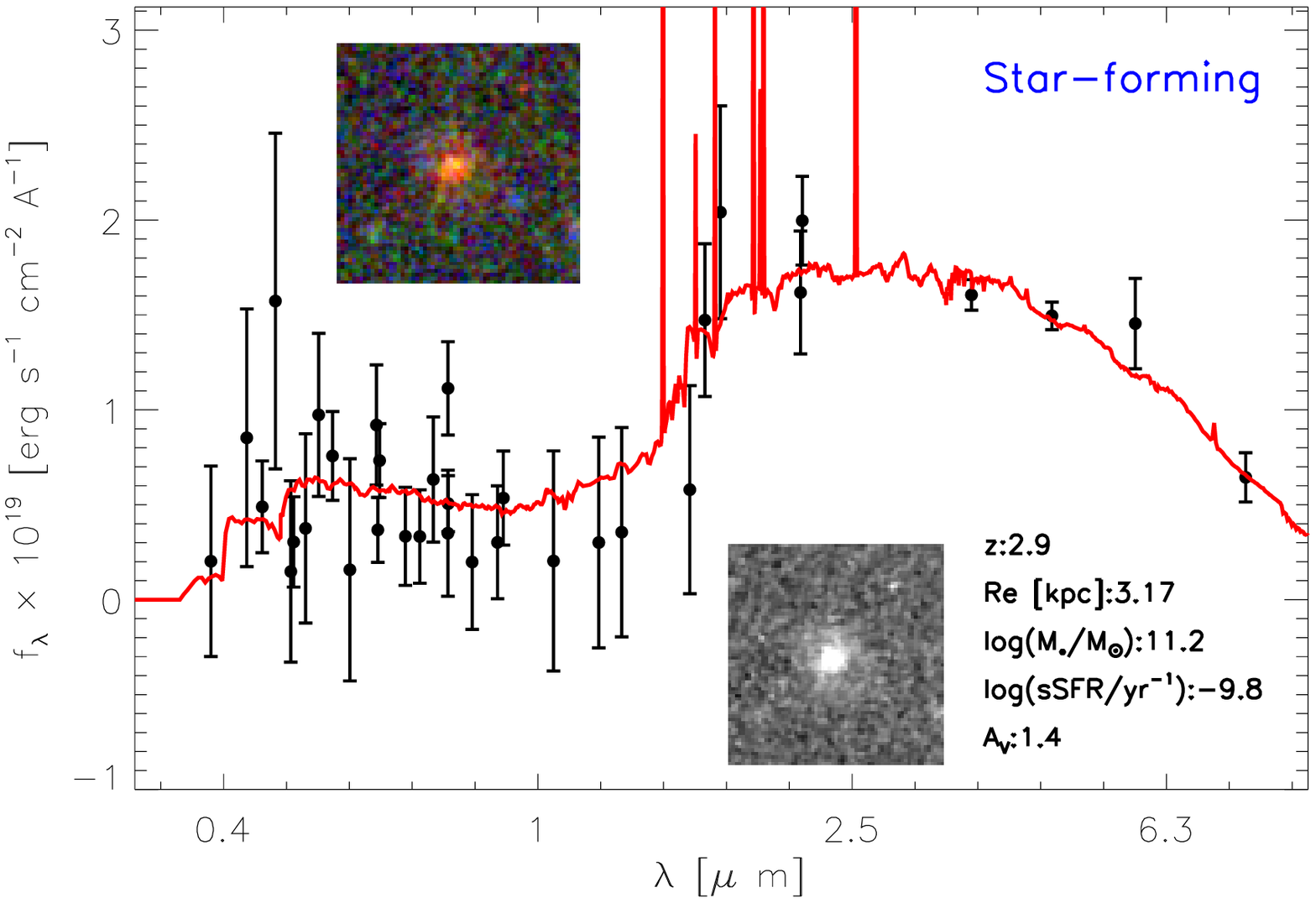} \\
\includegraphics[width=5.6cm]{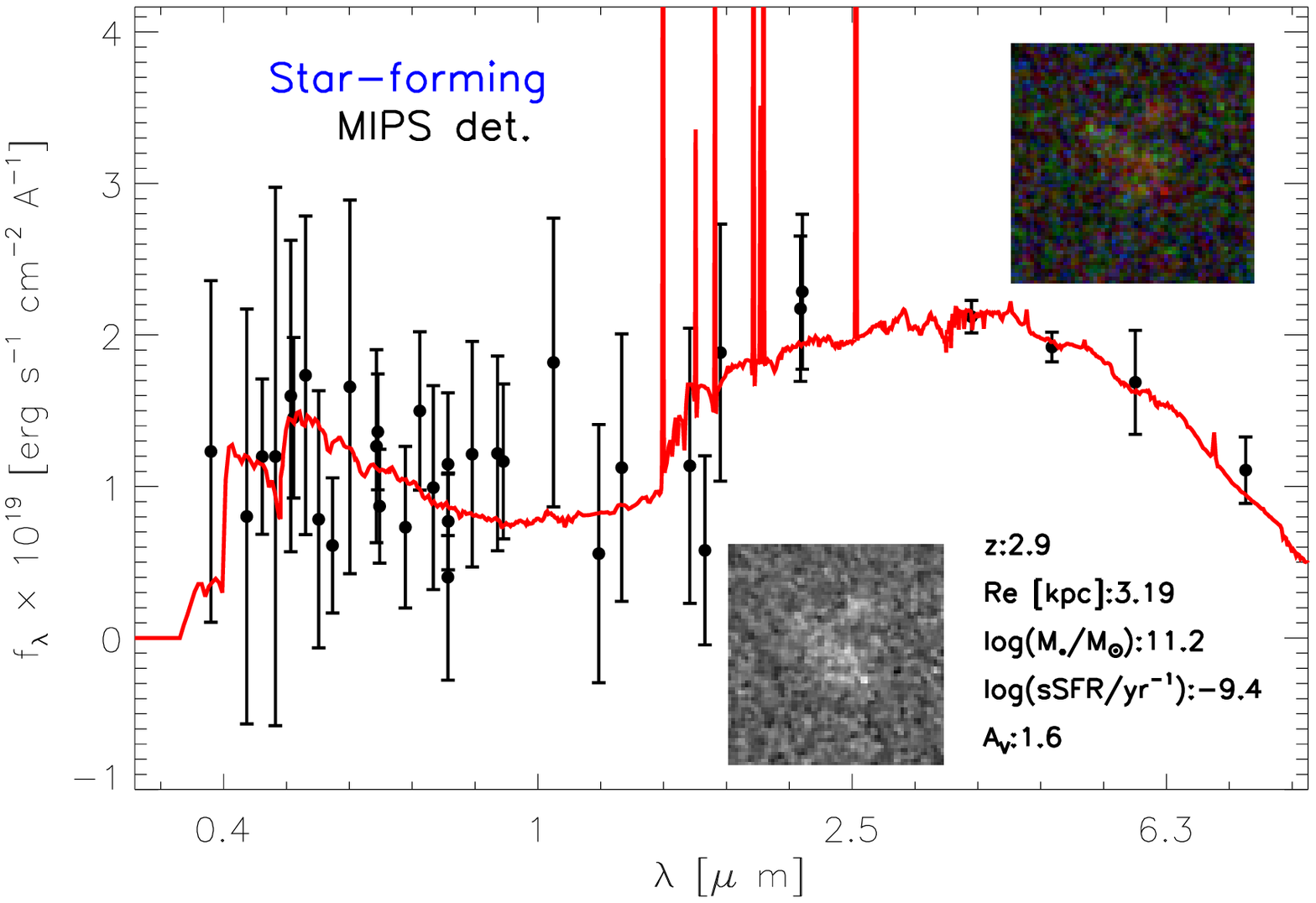} & \includegraphics[width=5.6cm]{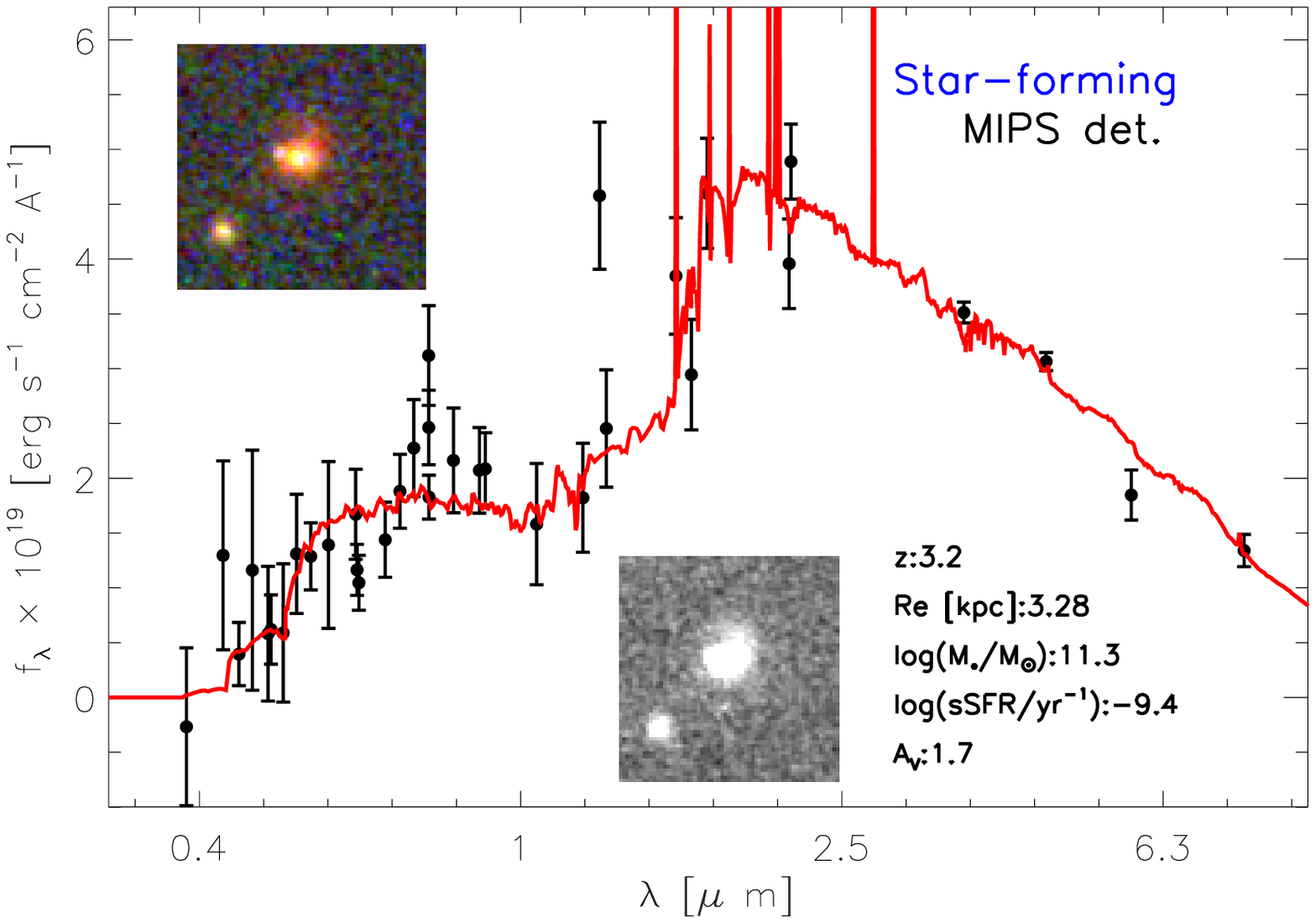} &  \includegraphics[width=5.6cm]{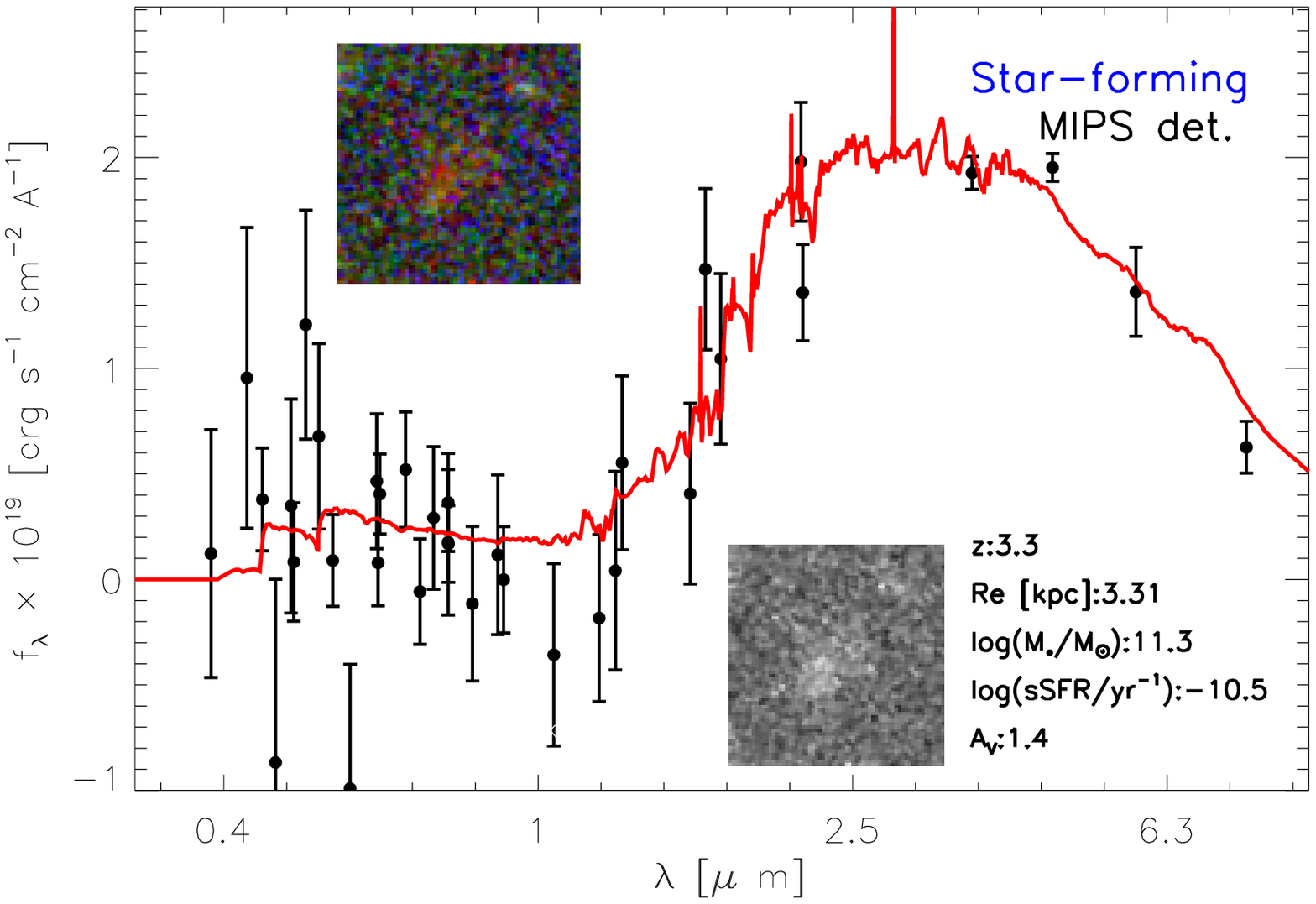} \\
\includegraphics[width=5.6cm]{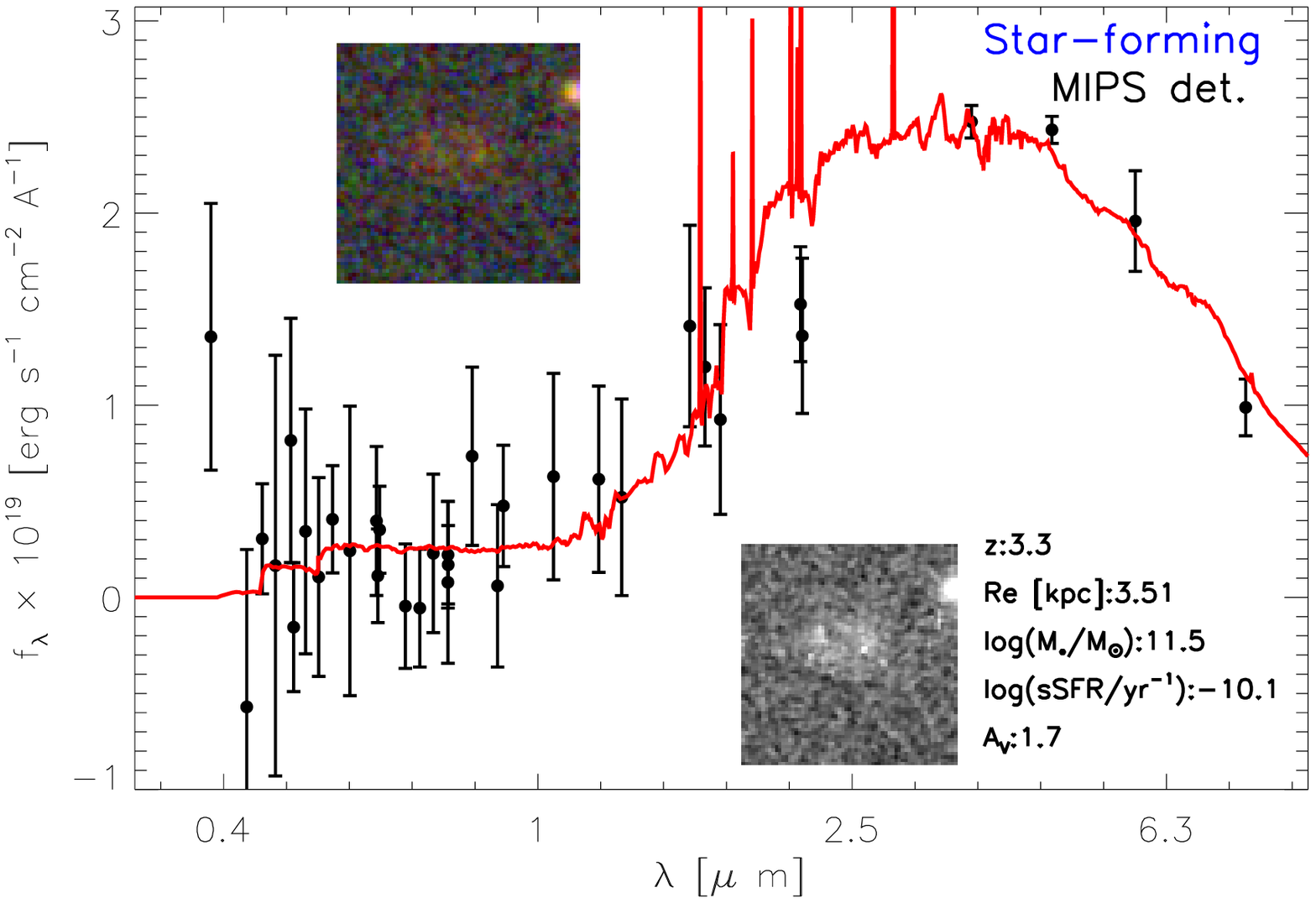} & & 
\end{tabular}
\caption{Observed SED from NMBS (black points) together with EAZY best-fitting SED (solid red curve) of our sample of galaxies with $\log(M_*M_\odot)>11$. A color cutout centered on the object and based on the F814W, F125W and F160W filters from CANDELS, with color scheme following \citet{lupton2004}, is presented in the upper part of each plot, while a cutout from the F160W filter is shown at the bottom of each plot is. The \emph{MIPS det.} label indicates those objects with a $3-\sigma$ MIPS detection (see text for details). The angular size of each cutout is $3.7"\times3.7"$.\label{fig:sed_cutout}}
\end{figure*}

In subsequent sections we will evolve our $z\sim3$ sample to $z\sim2$ to assess its correspondence to the $z\sim2$ observed population. In principle we should also estimate completeness levels in stellar mass for this evolved population. This could be done in a way similar to that implemented for the $z>3$ sample. In particular, this means that, for each object, the final completeness correction for the $z\simeq2$ sample would be the product between the completeness correction at $z\simeq3$ and that at $z\simeq2$. However, as shown by \citet{brammer2011}, the stellar mass completeness at $z\simeq2$ is 95\% or better for galaxies with stellar mass $\log(M_*)\gtrsim11$, which coincides with our stellar mass selection limit. This implies that no further correction is needed for the massive $z\simeq2$ sample.

\subsection{Cosmic Variance}

Given the small region of overlap between the NMBS-COSMOS and CANDELS-COSMOS fields, the statistical uncertainties associated with fluctuations of the large-scale density (i.e. the cosmic variance) are playing a non negligible role. We measured such effects following the recipe by \citet{moster2011}. A halo distribution model is used to relate the stellar mass to the dark matter halo as a function of redshift; the galaxy bias is then estimated via dissipationless N-body simulations. The cosmic variance is first computed on dark matter haloes, and then converted to galaxy cosmic variance by applying the galaxy bias. The average relative error due to cosmic variance for the $z>3$ massive sample is 35\%. This value was added in quadrature to the Poisson errors in the number densities we computed.

\subsection{The sample}

Our primary sample is composed by those galaxies brighter than $Ks=23.4$AB, whose redshift is compatible with $z>3$, with stellar mass $M_*>10^{11}M_\odot$, and which lie on the CANDELS-COSMOS frame. This selection yields 10 galaxies. Given that their associated completeness is $\sim70\%$, this is the only sample which allows us to perform a quantitatively robust analysis. Specifically, all number densities quoted in this work are based on the above sample. Their SEDs are shown in Figure \ref{fig:sed_cutout}, together with the F160W and color cutout.

\section{Results}

\label{sec:results}

\subsection{Massive $z>3$ galaxies}

The distribution of the specific star formation rate (sSFR=SFR/$M_*$) as a function of the effective radius $R_e$ for the $3<z<4$ sample is shown in Figure  \ref{fig:re-ssfr}. Points are color-coded according to their stellar mass. The effective radius from a Sersic profile fit to the point-source objects is $0.11\pm0.01$"; its projection to  $3<z<4$ corresponds to $\sim0.8$~kpc and is marked by the solid grey region, identifying a limit for the PSF-convolved Sersic profile. This marks an empirically determined limit below which we cannot robustly determine the size. Given our stellar mass completeness measurement, galaxies satisfying the 70\% completeness limit coincide with those whose stellar mass $M_*>10^{11}M_{\odot}$, identified by the red squares.

In the following, if not otherwise specified, we refer to \emph{massive} galaxies as those with stellar mass $M_*>10^{11}M_\odot$; we define \emph{compact} galaxies as those with $R_e<2$~kpc and \emph{quiescent} galaxies as those whose sSFR$<10^{-11}$ yr$^{-1}$. Our definition of quiescent galaxies is based on the work of \citet{kriek2006}, where a population of massive quiescent galaxies is spectroscopically identified at $z\sim2.3$. In their sample, the average sSFR is $4\times10^{-12}$~ yr$^{-1}$, with all galaxies having $\log(\mbox{sSFR/yr}^{-1})<-11$. For galaxies of $\log(M_*/M_\odot)=11$, this corresponds to SFR$<1M_\odot$ yr$^{-1}$. While galaxies with sSFR$>10^{-10}$~yr$^{-1}$ can be reasonably considered star-forming, those with intermediate sSFR ($-11<\log(\mbox{sSFR/yr}^{-1})<-10$) are galaxies with suppressed star-formation activities (relative to their stellar masses) with respect to the star-forming sequence observed at $z\sim2$ (e.g. \citealt{whitaker2012,szomoru2012}), although not necessarily quiescent. We therefore adopt a more conservative definition of  quiescent galaxy (i.e., sSFR$<10^{-11}$ yr$^{-1}$) in place of sSFR$<10^{-10}$~yr$^{-1}$, sometimes used in the literature.  We also checked Spitzer/MIPS 24 $\mu$m data relative to each massive galaxies. All but one star-forming galaxies are detected at more than 3$\sigma$. No significant MIPS detection is found for 2 of the 4 massive quiescent galaxies, consistently with their quiescent nature from SED modeling. However, two quiescent massive galaxies have significant MIPS fluxes ($\sim80$ and $\sim140\mu$Jy), which would imply large star formation rates (700-900 $M_\odot$yr$^{-1}$; \citealt{whitaker2012}) if the 24$\mu$m emission were associated to dust-enshrouded star formation. We note however that at $z>3$, the MIPS 24$\mu$m band samples rest-frame wavelengths shorter than 6$\mu$m, i.e., emission from hot dust. As such, the conversion from mid-IR fluxes to SFRs is very uncertain, especially at the high redshifts targeted in our work. Moreover, additional contamination due to the emission from the dusty torus of an AGN can potentially contaminate the MIR fluxes probed by the MIPS 24$\mu$m band. For these reasons, we will use the SFRs derived from the SED modeling throughout the paper, noting that half of the sub-sample of massive quiescent galaxies could actually be highly obscured, star-forming galaxies. Observations in the far-IR (e.g., ALMA) are needed to robustly quantify the level of obscured star formation and to confirm the quiescent nature of these galaxies.

The data show that at $z>3$, $M_*>10^{11}M_\odot$ galaxies that are quiescent tend to be compact, while those that are star-forming are more extended, similarly to the distribution found at $z\sim2$. Most of the galaxies ($\sim 70\%$ - see the inset in Fig. \ref{fig:re-ssfr}) are characterized by $R_e<2$ kpc, with a median of 1.5~kpc. However we caution that these values refer to the full sample which is not complete in stellar mass. The average size for the MQGs in the stellar mass complete sample is 1.2~kpc (0.6~kpc pc if the two galaxies with MIPS 24 $\mu$m detection are excluded); the corresponding value for the massive, star-forming sample is 3.1~kpc.

The plot also shows a lack of extended ($R_e>2$~kpc) quiescent galaxies, indicating that the massive quiescent  (elliptical) galaxies observed in the Local Universe were not yet completely formed when the Universe was $\lesssim2$~Gyr old. This could also be a surface brightness selection effect, although the width of NMBS PSF compared to CANDELS PSF causes all objects with $R_e<5$~kpc to be detected as point-sources; the selection effect should then act only on the very extended objects.

The plot in Figure \ref{fig:re-ssfr} shows that MQGs were already present at $z>3$, confirming previous results (see e.g. \citealt{marchesini2010} which first characterized the properties of a mass-complete sample of $3<z<4$ galaxies, finding both quiescent and star-forming galaxies). For $M_*>10^{11}M_\odot$ and sSFR $ < 10^{-11}$ yr$^{-1}$ we count four galaxies, corresponding to a completeness-corrected number density of $n_Q= 5.2_{-3.1}^{+4.6}\times10^{-6}$~Mpc$^{-3}$ ($n_Q=3.6_{-2.4}^{+4.0}\times10^{-6}$ Mpc$^{-3}$ if the two galaxies with MIPS 24 $\mu$m detection are excluded). The co-moving volume was computed assuming a redshift range $2.8<z<4.0$. Upper and lower error bars were computed following the recipe by \citet{gehrels1986}; cosmic variance was finally added in quadrature.

\begin{figure*}
\epsscale{1.0}
\begin{tabular}{cc}
\includegraphics[width=8.5cm]{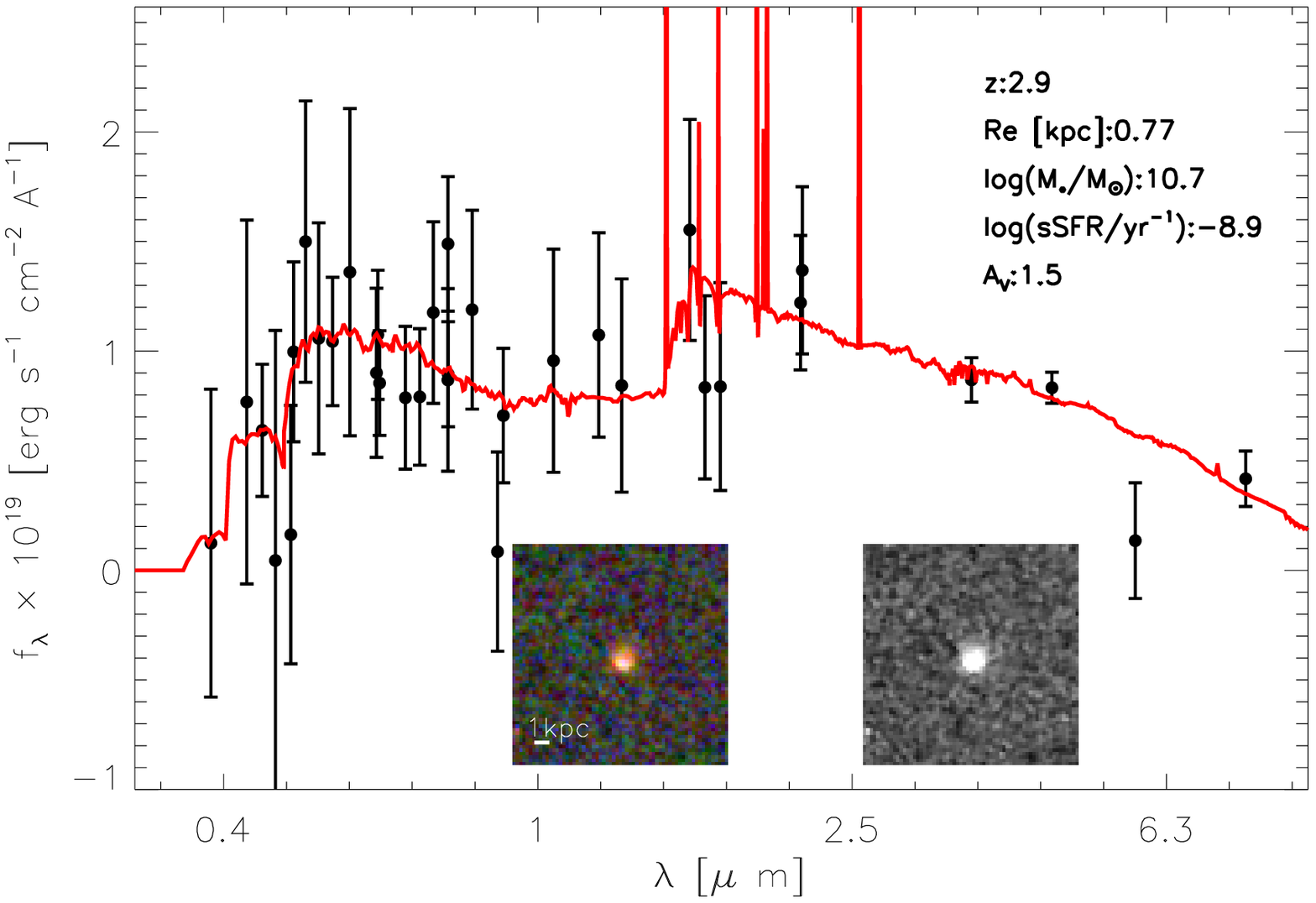} & \includegraphics[width=8.5cm]{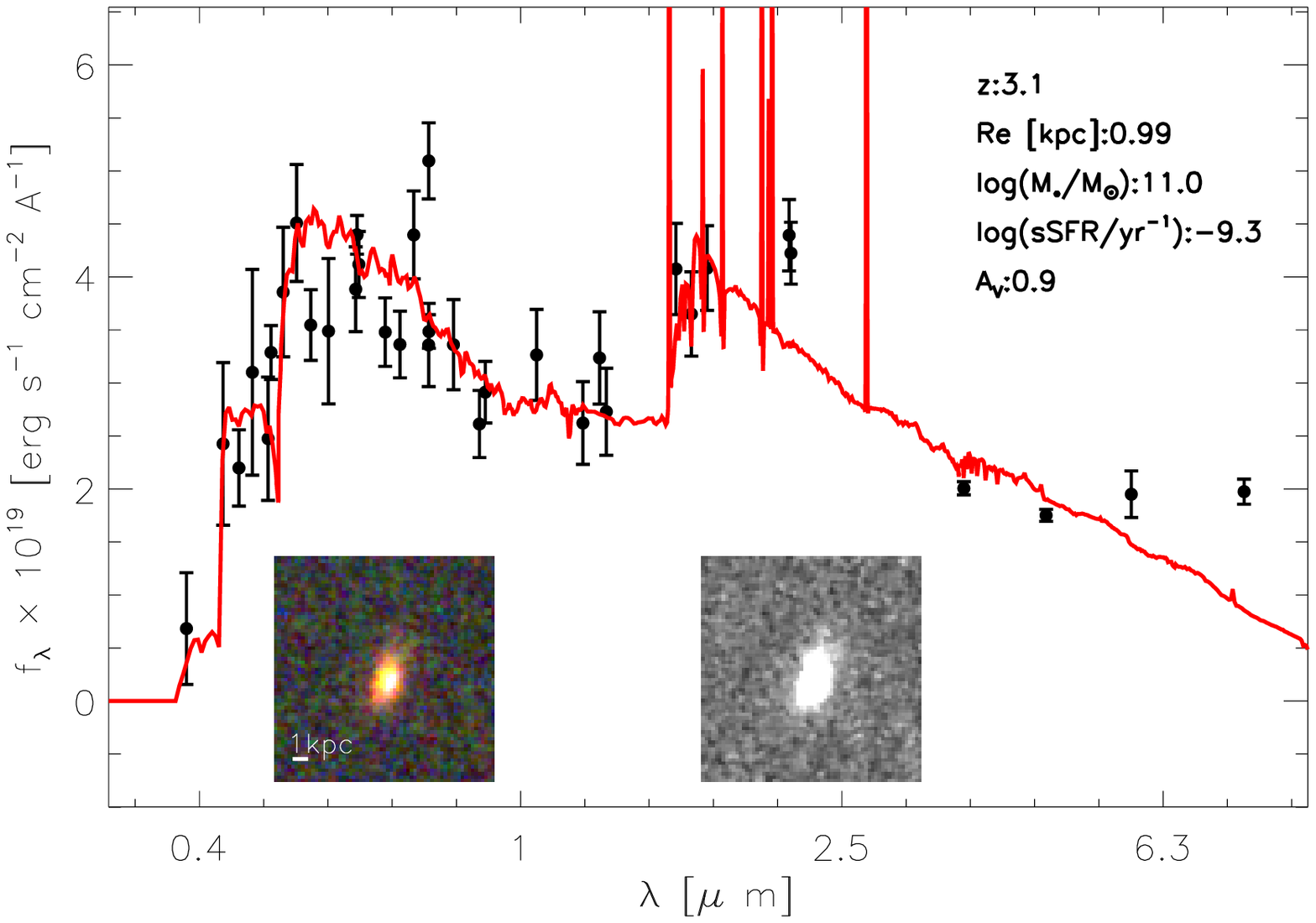} \\
\includegraphics[width=8.5cm]{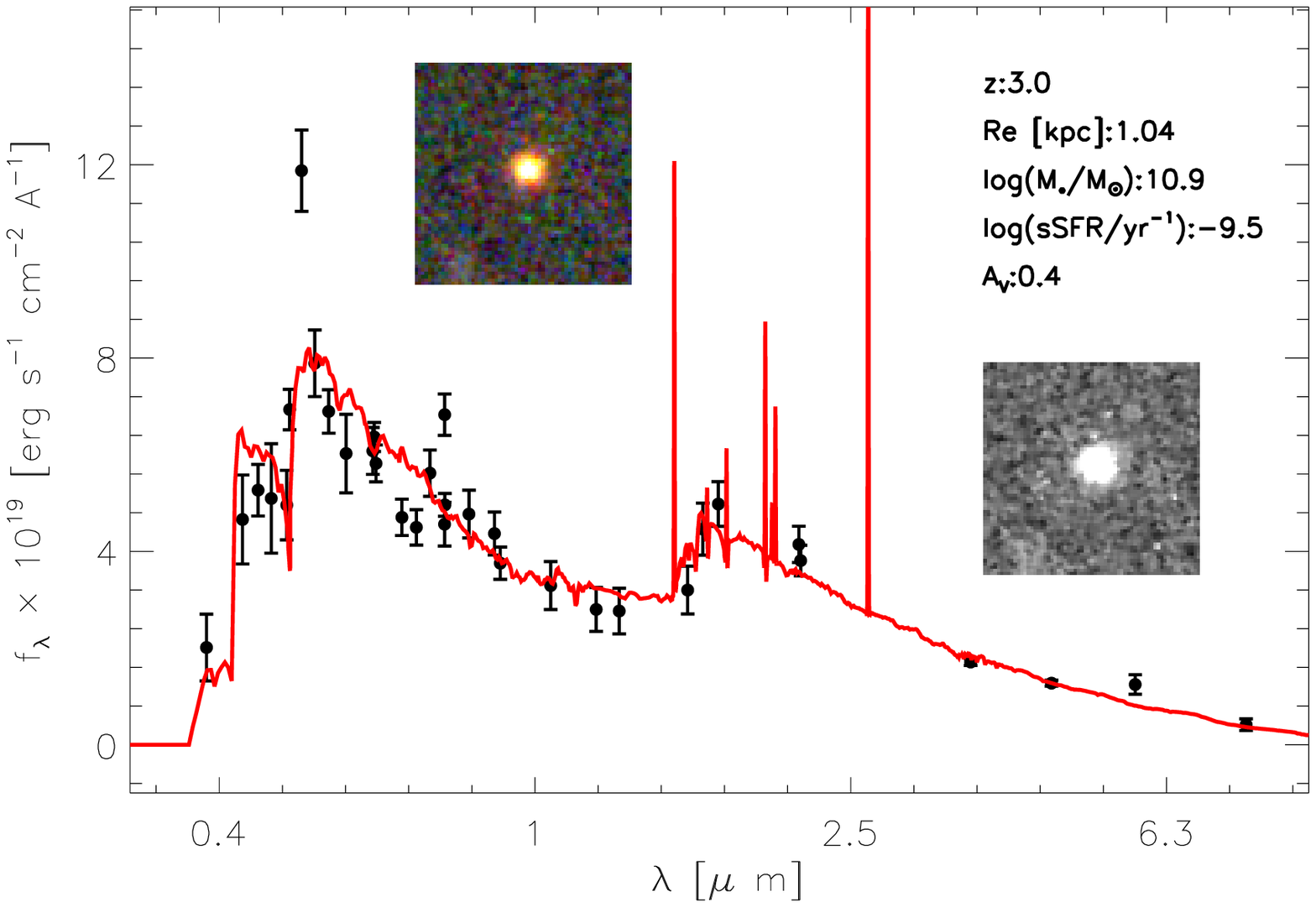} & \includegraphics[width=8.5cm]{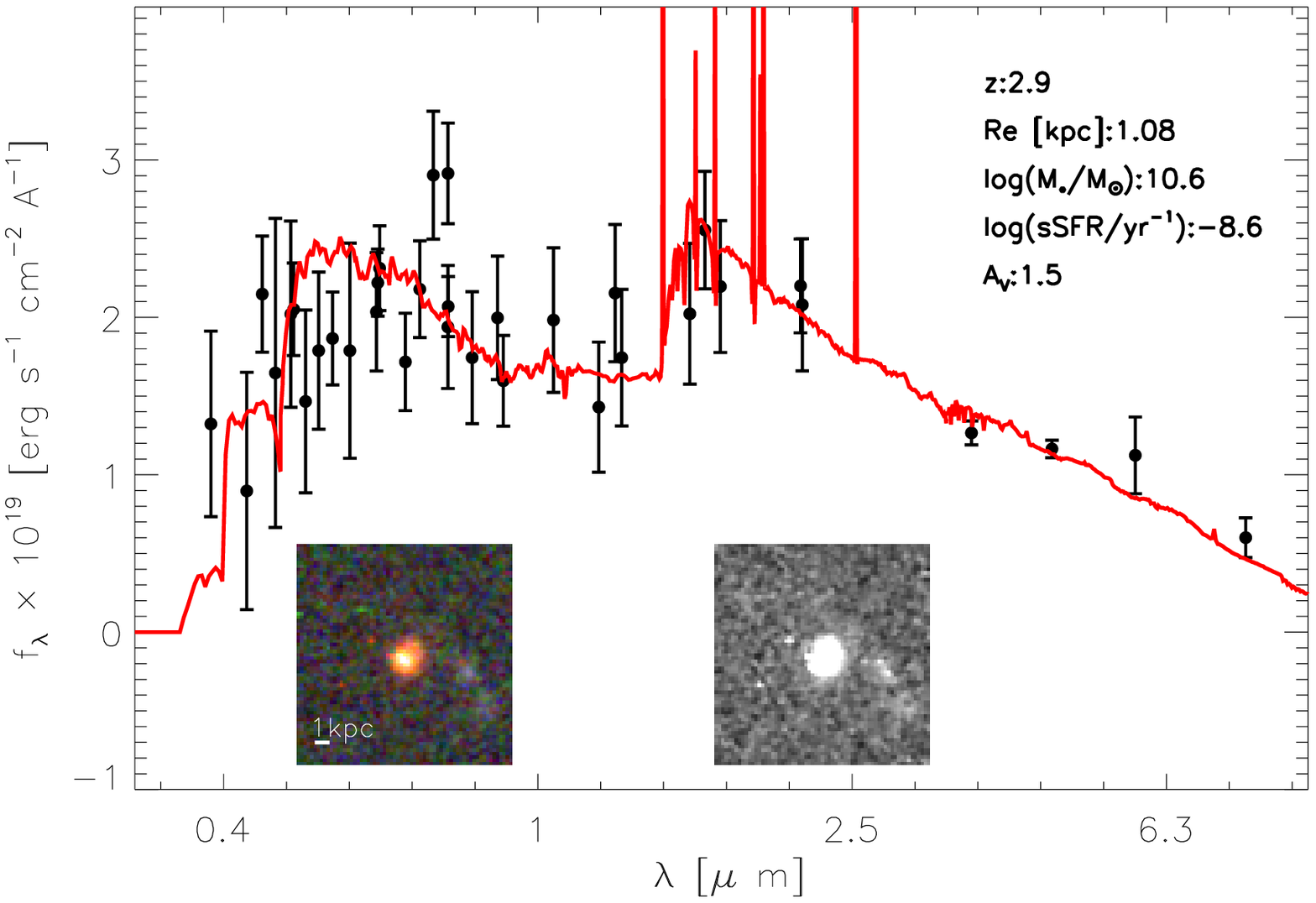} \\
\includegraphics[width=8.5cm]{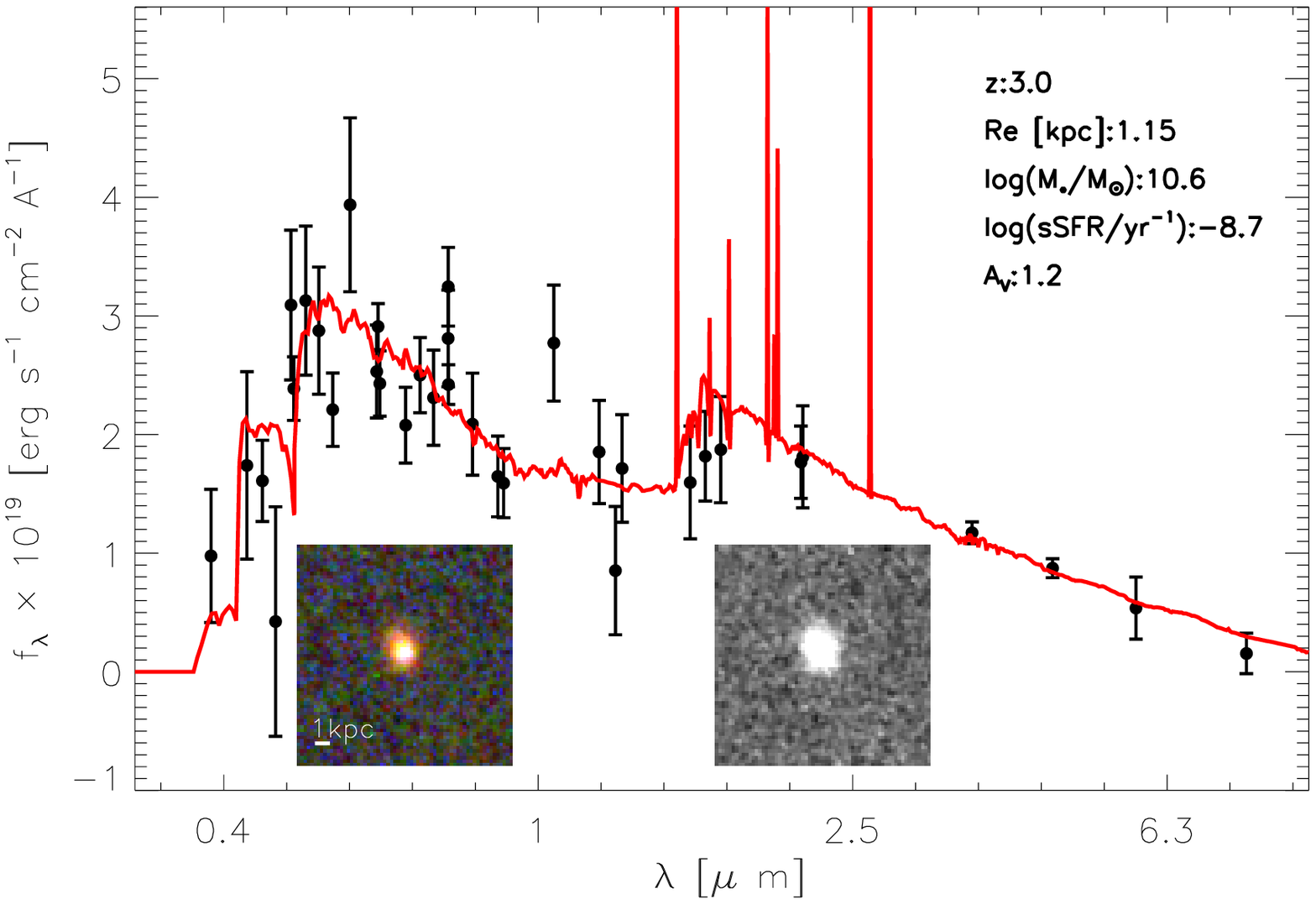} & 
\end{tabular}
\caption{Observed SED from NMBS (black points), EAZY best-fitting SED (solid red curve), color cutout built from F814W, F125W and F160W CANDELS frames and F160W cutout for our sample of compact, star-forming galaxies with $10.6<\log M_*/M_\odot<11.0$. The angular size of each cutout is $3.7"\times3.7"$.\label{fig:cmsf}}
\end{figure*}

Figure \ref{fig:re-ssfr} clearly shows that compact, MQGs were already present at $z>3$. This is one of the main results of this work. Studies so far have found this population of galaxies up to $z\sim2$ (see e.g. \citealt{kriek2006,bezanson2009}). Our data allows us to push back in time the appearance of compact, massive, quiescent galaxies from when the Universe was $\approx 3$ Gyr old to an age of less than $2$~Gyr.

\begin{figure}
\epsscale{1.0}
\plotone{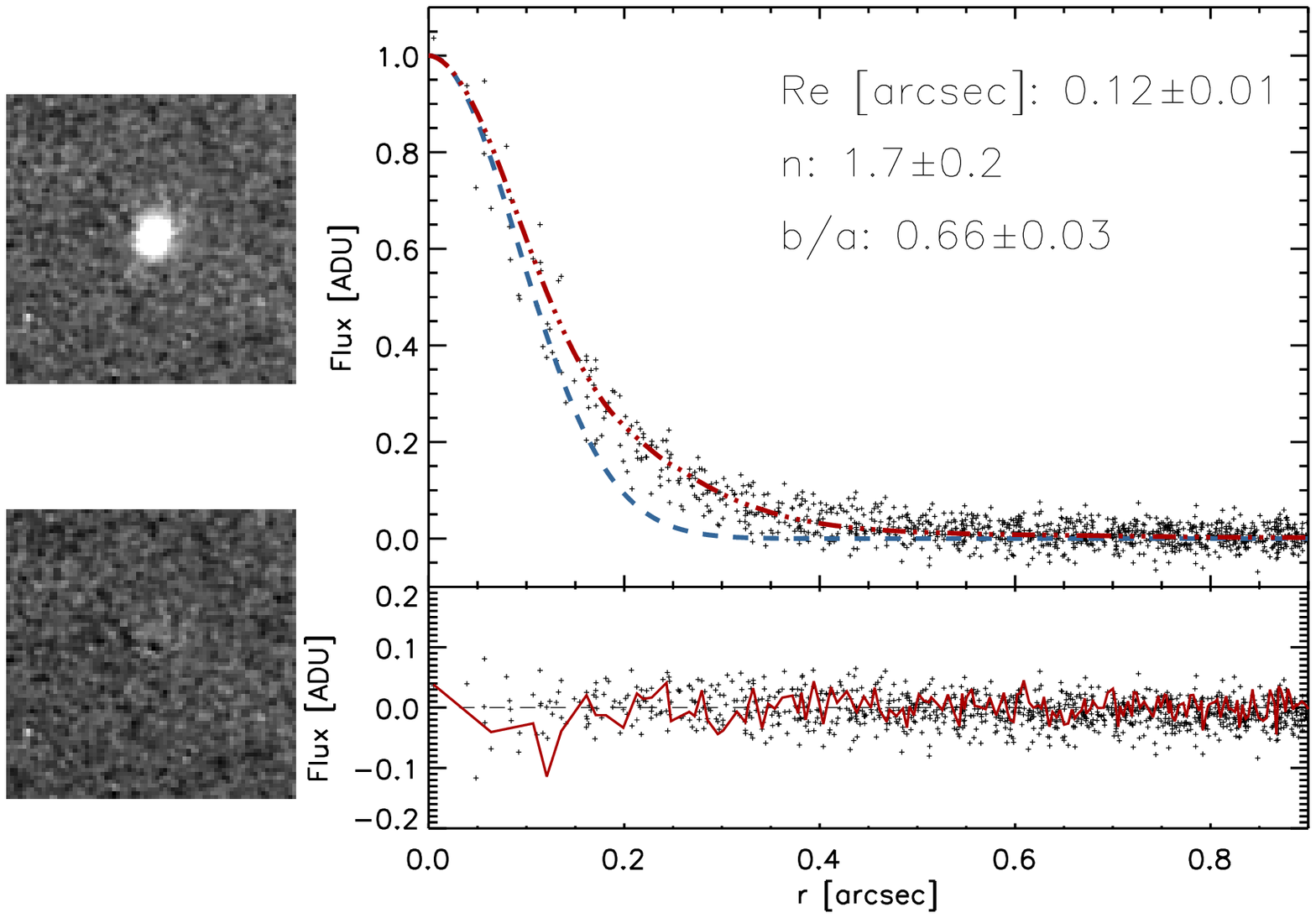}
\caption{The panels on the left show the stacked image for three out of the five compact star-forming galaxies with high stellar mass at $z>3$, obtained after excluding the two galaxies with very different aspect ratio and position angle  (top) and the residual image resulting from the fit (bottom). The panels on the right show on top the circularized radial profile (arbitrary units) of the stacked image indicated by the black points. The triple-dotted dashed red curve marks the PSF-convolved Sersic profile recovered by GALFIT, while the blue dashed line represents the profile of the PSF. The bottom panel shows the residuals measured directly from the GALFIT residual image (black points) and smoothed difference between the PSF-convolved analytic profile and the original input image (solid red line). The light profile closely resembles that of the PSF and it does not show evidence for a faint extended halo.\label{fig:stacked}}
\end{figure}

 \begin{figure}
\hspace{-0.7cm}
\includegraphics[width=9.5cm]{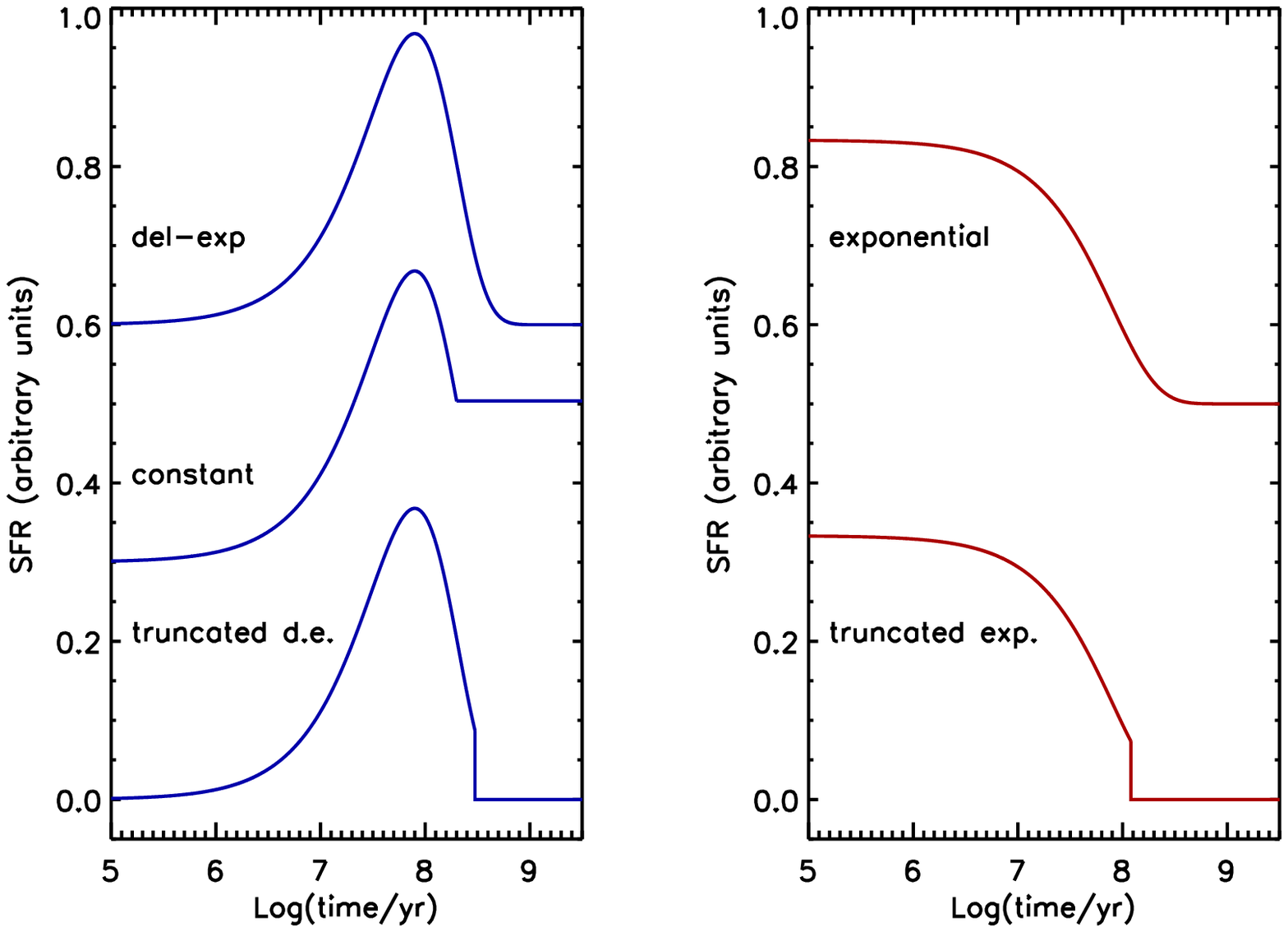}
\caption{Examples of the SFHs resulting from the different evolution scenarios assumed in this work. \emph{Left panel:} From top to bottom, are the delayed-exponential SFH, the constant SFH and the truncated delayed-exponential SFH. \emph{Right panel:} from top to bottom, the exponential SFH and the truncated exponential SFH. The adopted parameters are representative of the average $z>3$ population for all but the truncated-exponential SFH: e-folding time ($\tau$) of $8\times10^{7}$~yr, and age of $3\times10^{8}$~yr. The age of the truncated-exponential SFH corresponds to $2\times10^{7}$~yr. Arbitrary normalization and offset were applied in order to increase readability.\label{fig:sfh_time}}
\end{figure}

As we will show in Sec. \ref{sec:cmq}, these objects do not exist in significant enough numbers to explain the abundance of compact, massive, quiescent galaxies (CMQGs) at $z\sim2$.

\subsection{Compact, star-forming galaxies at $z\sim3$ with large stellar masses}
 
\label{sect:cmsf} 
 
 The existence of a relation between $M_*$ and $R_e$ (see e.g. \citealt{shen2003,mosleh2011}) implies that we can not consider a single value for the $R_e$ when selecting compact galaxies at different stellar mass ranges. Assuming $R_e \propto M_*^{0.32}$, valid for galaxies at $2.5<z<3.5$, \citep{mosleh2011}, an $R_e=2.0$~kpc for a $\log (M_*/M_\odot)=11$ galaxy scales to $R_e=1.4$~kpc for $\log (M_*/M_\odot)=10.5$ and  $R_e=1.0$~kpc for $\log (M_*/M_\odot)=10.0$. Figure \ref{fig:re-ssfr}  then shows that there are a number of compact star-forming galaxies, spanning the whole range of stellar masses. In particular, and more interestingly, in our sample there are 5 galaxies with high sSFR ($\log({\rm sSFR/yr}^{-1})\sim-9$), high stellar masses ($10.6<\log(M_{*}/M_\odot)<11)$ and with $R_e<1.4$~kpc (i.e. they are compact). Their redshift is above $z=2.6$ with 68\% confidence level, excluding contamination from low redshift galaxies with broad probability distributions of photometric redshifts. From Figure \ref{fig:compl}, the stellar mass completeness of $\log(M_*/M_\odot)\sim10.6$ galaxies is lower than 50\% for $M/L>0.9M_\odot/L_{\odot,V}$, i.e. we could possibly be missing star-forming galaxies with significant dust absorption. In Figure \ref{fig:cmsf} we present their SEDs along with F160W and color cutouts.  We note however, that one out of the 5 objects has both 5 and 8~$\mu$m excess that could be the imprint of a Type 2 AGN. However, there is no evidence for this AGN in the rest-frame optical SED and we therefore do not think that it is causing a significant bias in our sizes, although obscured AGN are a potential source of uncertainty.

To determine if there is a faint extended component in the compact star-forming galaxies and to better assess our size measurements, we directly stacked the images for the 5 objects.  We normalized each image tile to its peak flux to prevent any single object from dominating the stack. The brightness profile of the stacked image was analyzed using GALFIT and a S\'ersic profile, obtaining $R_e=0.53"\pm0.02"$ and $n=3.5\pm0.2$. Assuming an average redshift of $z\sim3$, this corresponds to $R_e\sim7.7$~kpc. However, we note that visual inspection of the images of the individual objects shows that the aspect ratio and position angle of two galaxies are very different from the other three, causing significant broadening of the profile. We therefore repeated the stacking excluding these two sources, obtaining $R_e=0.12"\pm0.01"$ ($0.9$~kpc at $z\sim3$) and $n=1.7 \pm 0.2$, further supporting their compact configuration. The stacked image and its surface brightness profile for this second analysis are shown in Figure \ref{fig:stacked}. 

These compact, star-forming galaxies  with high stellar mass are likely candidates for being the progenitors of the compact, MQGs observed at $z\simeq2$. However, as we will show in the next sections, the number density associated to this sample of objects cannot explain by itself the number densities observed for the compact, MQGs at $z\sim2$.

\section{What were the progenitors of the $z\sim2.3$ compact, massive, and quiescent galaxies?}

\label{sec:learn}

\subsection{Evolving $z>3$ galaxies to $z=2.3$}

In this section we describe the evolution of the $z>3$ population of  massive ($M_*>10^{11} M_\odot$)  galaxies down to $z=2.3$, paying specific attention to the progenitor population of the massive, quiescent  (${\rm sSFR}<10^{-11} $ yr$^{-1}$) galaxies at $z\sim2.3$ \citep{kriek2006,vandokkum2008,bezanson2009}. We used the stellar masses and SFRs from the public NMBS catalog. These were computed using the FAST code \citep{kriek2009}, adopting a \citet{kroupa2001} IMF, a delayed-exponential star formation history (SFH) and solar metallicity. 

The stellar mass and the SFR of each $z>3$ galaxy was evolved to $z=2.3$ using the GALAXEV program from the \citet{bruzual2003} SSP models using a set of different SFHs. Specifically we adopted 5 distinct SFHs: a constant SFR (CSF), an exponentially declining SFR (E), an exponentially declining with quenching of star formation 100Myr after the observed redshift SFR (TE), a delayed-exponential SFR (DE), and its quenched version 100Myr after the observed redshift (TDE). We allowed each SFH to progress starting at the mass and SFR given by FAST. The adopted SFH are schematically presented in Figure \ref{fig:sfh_time}.

\begin{figure*}
\epsscale{1.0}
\begin{tabular}{cc}
\includegraphics[width=8.5cm]{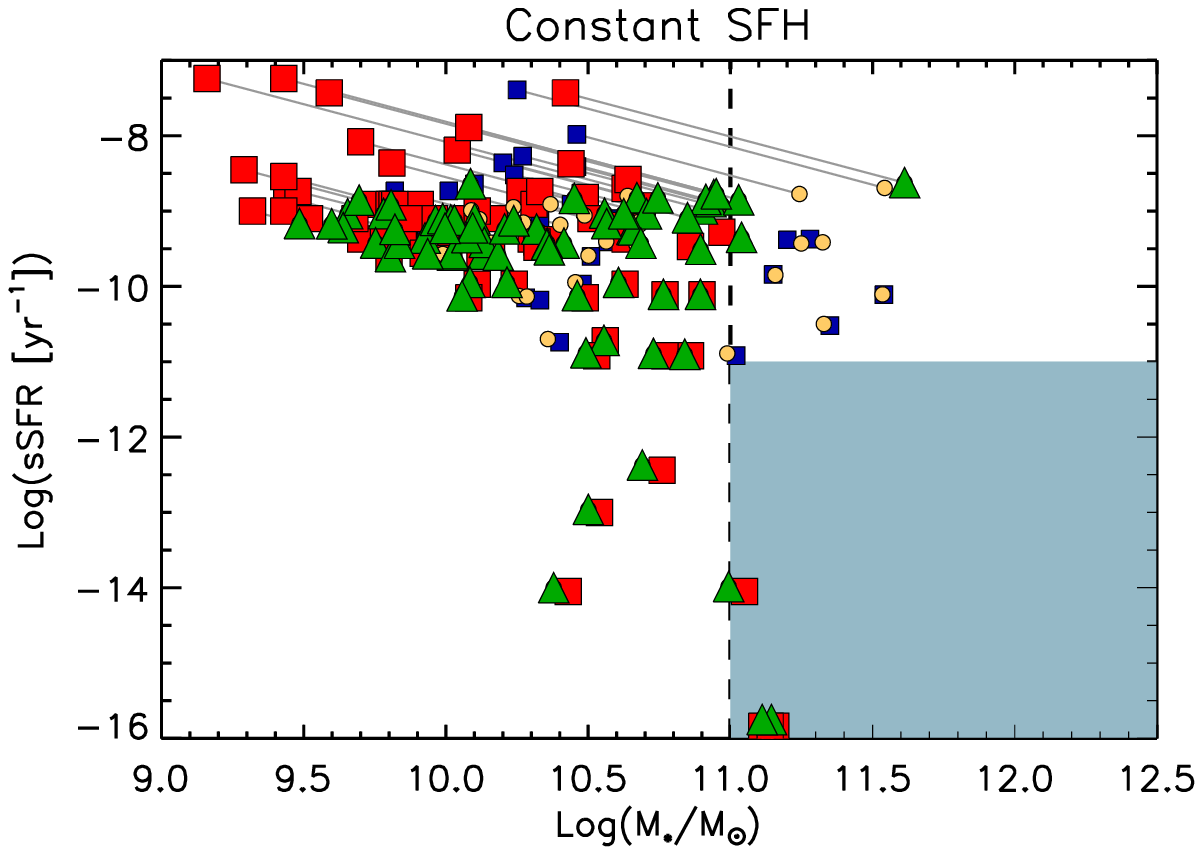} & \includegraphics[width=8.5cm]{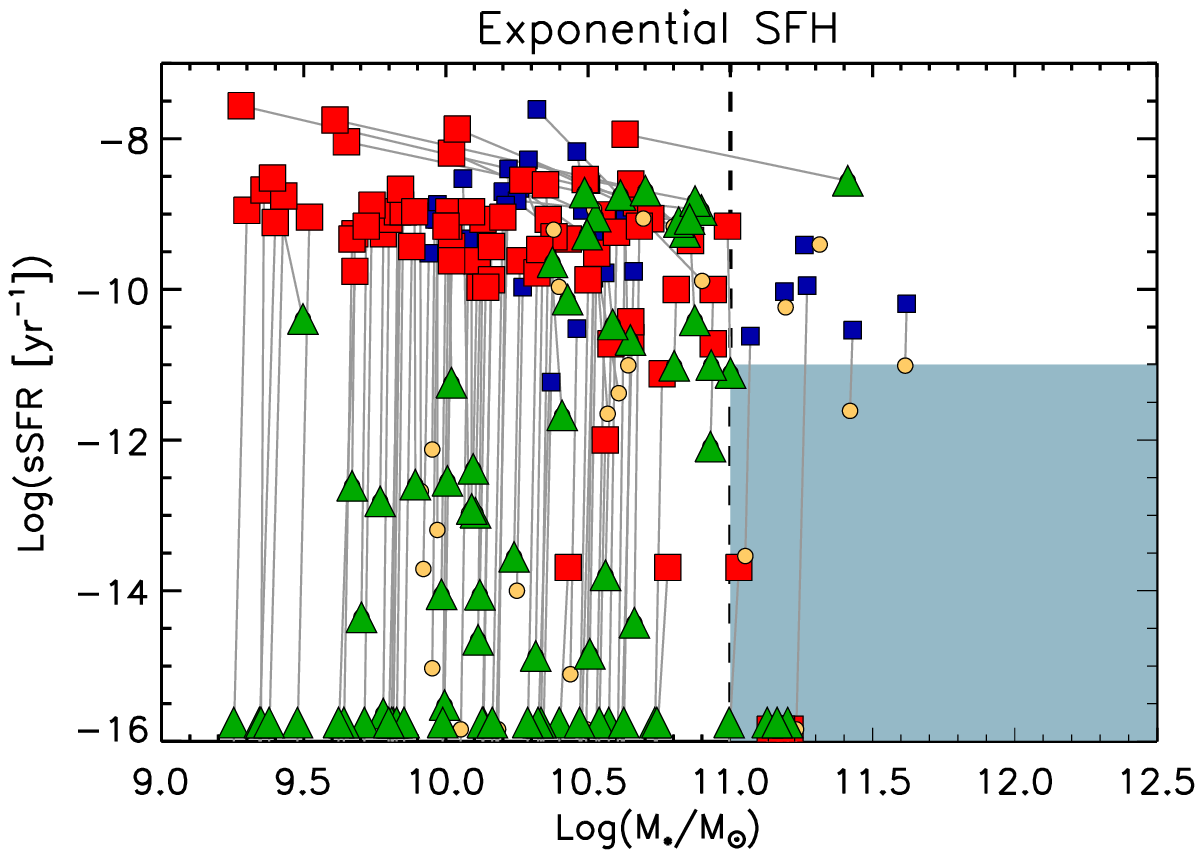} \\
\includegraphics[width=8.5cm]{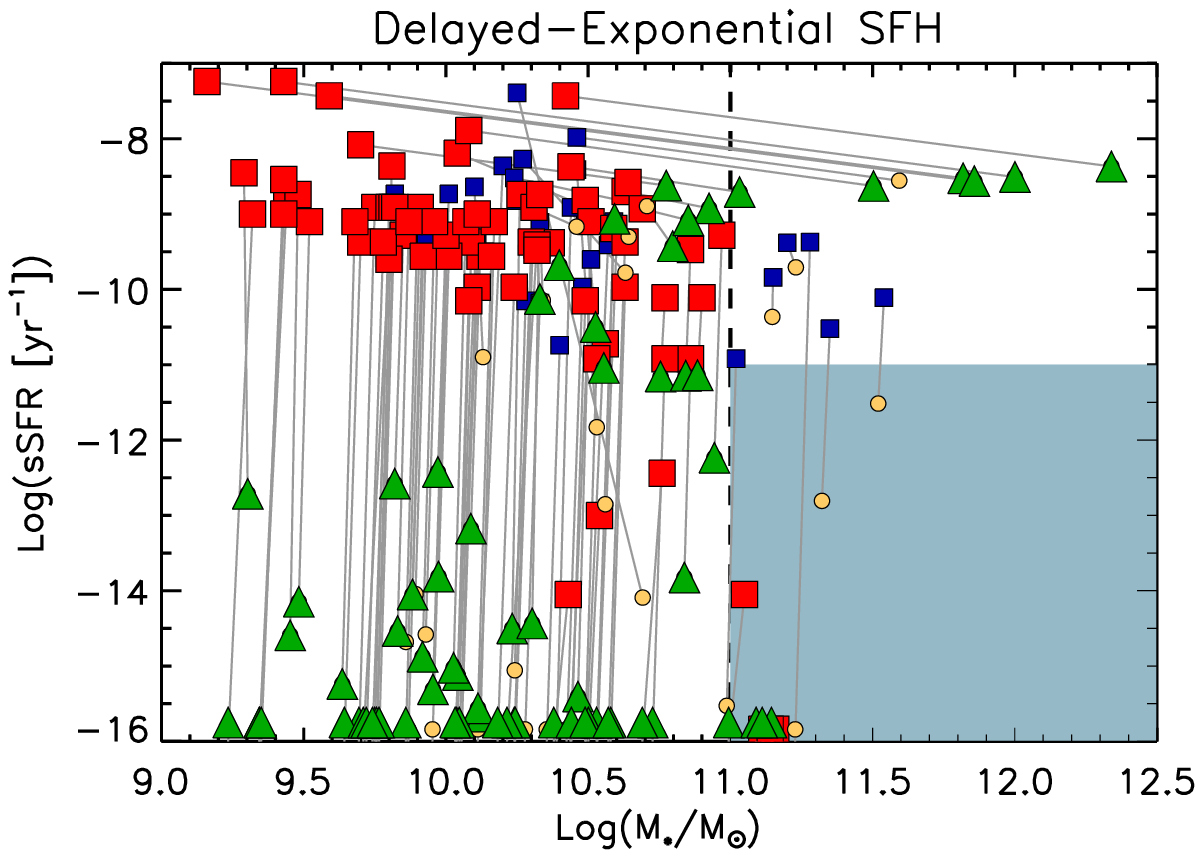} & \includegraphics[width=8.5cm]{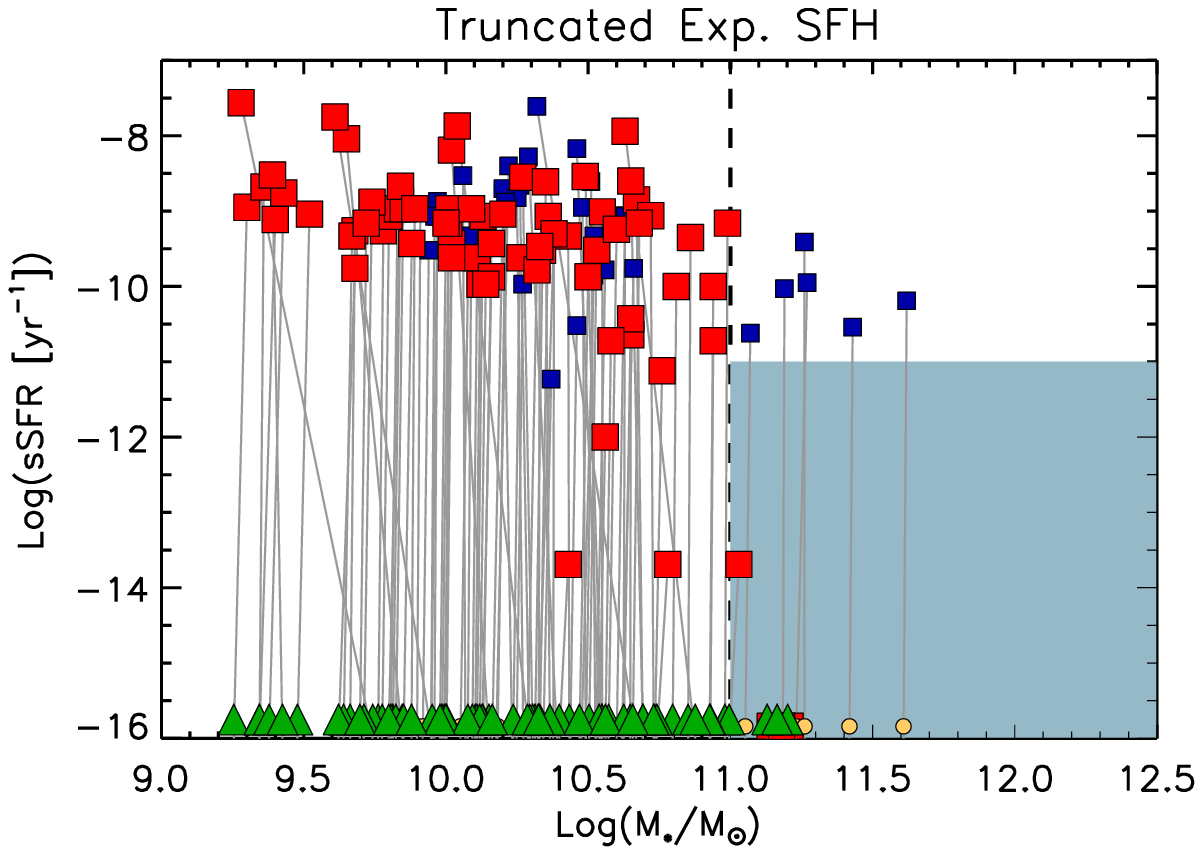} 
\end{tabular}
\caption{Evolution of the $z>3$ galaxy sample in the  sSFR-$M_*$ plane to $z=2.3$ for 4 different SFHs (left to right, top to bottom: constant, exponential, delayed-exponential and truncated exponential SFHs). Red big squares mark compact  (i.e. $R_e<2$~kpc) $z>3$ galaxies, while blue smaller squares indicate the remaining $z>3$ galaxies. The position of each galaxy after evolution is indicated by the green triangles and yellow circles for descendants of compact and extended galaxies, respectively. The filled area marks the selection criteria adopted for the massive ($M_*>10^{11}$M$_\odot$) and quiescent (${\rm sSFR}<10^{-11}$ yr$^{-1}$) sample at $z=2.3$. The vertical dashed line indicates the stellar mass limit corresponding to our 70\% completeness. Barring significant size evolution, the descendants of the red squares that fall within the shaded box tell us the number of predicted $z\sim2$ compact, massive, quiescent galaxies.  The number of galaxies in the shaded box is possibly a lower limit as they are the result of evolving the full $z>3$ sample and not only those galaxies from the stella-mass complete sample. In particular, there may be galaxies with stellar masses below our completeness limit at $z\sim3$ which could nonetheless grow above the limit by $z\sim2.3$. The largest contribution to the change in number density in the evolution of the population of massive quiescent galaxies comes mainly from the decreased value of the SFR at $z=2.3$, rather than from an increase in stellar mass with cosmic time.\label{fig:SFHs}}
\end{figure*}

The $z>3$ sample and its evolution at $z=2.3$ are shown in the sSFR-$M_*$ plane in the four panels of Figure \ref{fig:SFHs} (we omit the truncated delayed-exponential case as its results resemble those from the truncated exponential SFH).

\label{sec:cmsf}

The panels in Figure \ref{fig:SFHs} show that galaxies with sSFR$<10^{-9}$yr$^{-1}$ do not significantly increase their stellar mass with time due to their low SFRs, but instead they keep approximately the same value, or slightly decrease it, due to the return of stellar mass to the gas phase in the ISM. The largest contribution to the change in number density in the evolution of the population of MQGs comes mainly from the decreased value of the SFR at $z=2.3$, rather than from an increase in stellar mass with cosmic time.\\

\begin{figure*}[t!]
\epsscale{0.47}
\begin{tabular}{cc}
\hspace{-0.7cm}
\includegraphics[width=9.4cm]{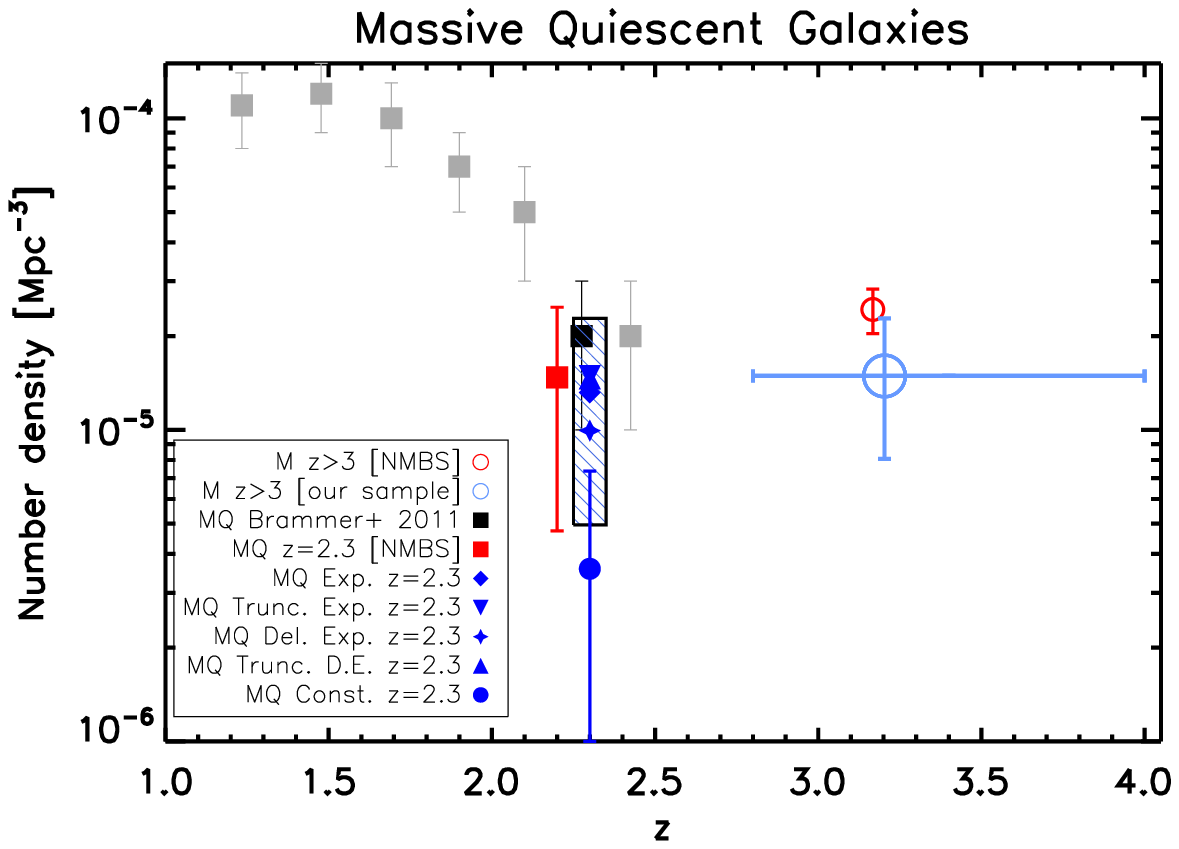} & \hspace{-0.7cm} \includegraphics[width=9.4cm]{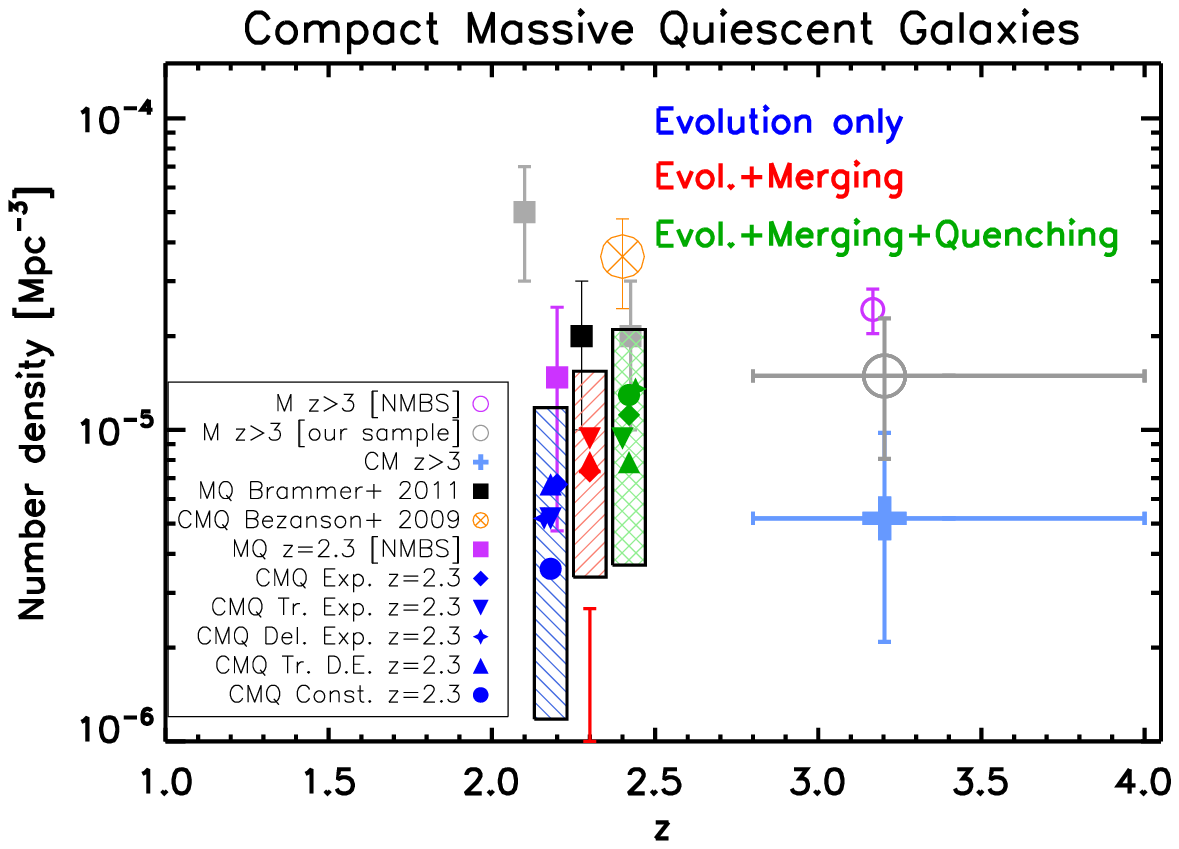}
\end{tabular}
\caption{\emph{Left panel}: number density of the massive $z>3$ sample with overlap with the CANDELS field (blue open circle), and computed from the full NMBS catalog (red open circle). The filled red square marks the number density of $z\sim2.3$ massive quiescent galaxies computed using the full NMBS catalog. We note that, while \citet{brammer2011} number density refers to quiescent galaxies selected via the UVJ diagram, our measurement of the $z\sim2.3$ massive quiescent population is obtained applying a sSFR$<10^{-11}$yr$^{-1}$ cut. The blue hashed box at $z\sim2.3$ encompasses the range in number density of massive quiescent galaxies as predicted by four of our SFH scenarios  (DE, TDE, E, and TE) for evolving all galaxies from $z>3$ to lower redshift.  The vertical extent of the bar also takes into account the expected cosmic variance. It is clear from this that there are enough massive galaxies at $z>3$ to account for the full massive quiescent galaxy population at $z\sim2$.  The blue filled circle at $z=2.3$ represents the number density and total error from the CSF model. The total number density of $\log(M_*/M_\odot)>11$ galaxies at $z>3$ is consistent with the population of massive quiescent galaxies at $z\sim2$; all except the CSFH are plausible SFHs in reproducing the evolution to $z\sim2$. \emph{Right panel}: number density of the original massive $z>3$ sample (open grey and magenta circles for the measurement obtained using our sample and the full NMBS catalog, respectively) and of the compact massive $z>3$ sample (thick blue plus), which encompasses the same objects as the $z>3$ MQ sample, and hence has the same number density. The colored boxes at $z\approx2.3$ mark the number densities of evolved compact, massive, quiescent galaxies under different scenarios.  The blue box represents pure evolution in the SFHs with no merging.  The red box and points represent the evolution after galaxies have been allowed to merge once randomly.  The green box includes the effect of the SFH, a random merger, and a post-merger truncation of the SFH. The blue and the green bars and the associated points have been shifted by an arbitrary value in $z$ to improve readability and should be considered at $z=2.3$. \citet{brammer2011} data is plotted only for $z>2$, i.e. where the vast majority of massive quiescent galaxies are also compact. Also plotted is the number density of $2.2<z<2.4$ massive, quiescent galaxies from the full NMBS catalog (red filled square). The number density of the compact, massive, quiescent population of galaxies obtained with pure SFH evolution to $z=2.3$ is  only marginally consistent with previous determinations; a simple merging model produces number densities in better agreement with previous determinations. Agreement at the $1-\sigma$ level is found with the number density of $z\sim2$ massive, quiescent galaxies computed from the full NMBS data set for the simple evolution scenario; the simple merging model increases the level of agreement of the number densities. We caution however that the number densities of the evolved samples are perhaps lower limits since they are the result of evolving all galaxies at $z>3$ and not just those for which we are mass complete. Specifically, there may be galaxies below our completeness limit at $z\sim3$ which could nonetheless grow above the limit by $z\sim2.3$.
 \label{fig:numden}}
\end{figure*}

In the following section, we present the number densities of massive, quiescent galaxies; the number densities of  the compact, massive, quiescent population is discussed in Section \ref{sec:cmq}. As a further and more consistent source of comparison, the number density of $z>3$ massive galaxies and of $z=2.3$ massive, quiescent galaxies was also computed using the full NMBS catalog. These two populations were selected  according to the same limits in stellar mass and SFR used for the other samples, i.e. $\log(M_*/M_\odot)>11$ and $\log$(sSFR/yr$^{-1}$)$<-11$. The full list of the number densities computed for the different choices of SFH can be found in Table \ref{tab:numden}.  The co-moving volume was computed assuming a redshift range $2.8<z<4.0$. We used upper and lower error bars from \citet{gehrels1986} and added in quadrature the cosmic variance. Figure \ref{fig:m_star-z} shows that there are 6 objects with $z<2.8$, although all of them have stellar masses $M_*<10^{11}M_\odot$. We repeated our analysis excluding the 6 objects with $z<2.8$ and obtained results quantitatively similar to those presented here.

\begin{table*}
\begin{center}
\caption{Number densities for the galaxy populations and SFHs adopted in this work.\label{tab:numden}}
\begin{tabular}{ccccccc}
\tableline
\tableline
Redshift & Population$^{(a)}$ & \multicolumn{5}{c}{Number density$^{(b)}$ ($10^{-6}$ Mpc$^{-3}$)}\\
\tableline
$z>3$ & M & \multicolumn{5}{c}{$14.9_{-6.8}^{+7.9}$} \\
	& M (Full NMBS) & \multicolumn{5}{c}{$24.3_{-3.3}^{+3.5}$}  \\
	& MQ & \multicolumn{5}{c}{$5.2_{-3.1}^{+4.6}$}  \\
	& & & & & & \\
Redshift & Population & \multicolumn{5}{c}{Number density ($10^{-6}$ Mpc$^{-3}$)} \\
\tableline
$z=2.3$	& MQ (Full NMBS) & \multicolumn{5}{c}{$14.7_{-10.5}^{+12.1}$}  \\
	& & & & & & \\
Redshift & Population & E ($10^{-6}$ Mpc$^{-3}$) & TE ($10^{-6}$ Mpc$^{-3}$) & DE ($10^{-6}$ Mpc$^{-3}$)& TDE ($10^{-6}$ Mpc$^{-3}$) & CSF ($10^{-6}$ Mpc$^{-3}$) \\
\tableline
$z=2.3$ & MQ &$13.2_{-6.2}^{+7.3}$ & $14.9_{-6.8}^{+7.9}$ & $9.9_{-5.0}^{+6.2}$ & $14.6_{-6.7}^{+7.8}$ & $3.6_{-2.4}^{+4.0}$ \\
	& CMQ & $6.7_{-3.7}^{+5.1}$ & $5.2_{-3.1}^{+4.6}$ & $5.2_{-3.1}^{+4.6}$ & $6.7_{-3.7}^{+5.1}$ & $3.6_{-2.4}^{+4.0}$ \\
	& & & & & & \\
\multicolumn{7}{c}{Merging} \\
$z=2.3$ & CMQ & $6.7_{-3.7}^{+5.1}$ & $9.8_{-4.9}^{+6.2}$ & $7.0_{-3.8}^{+5.2}$ & $9.1_{-4.7}^{+5.9}$ & $0.2_{-0.2}^{+2.6}$ \\
	& & & & & & \\
\multicolumn{7}{c}{Merging+Quenching} \\
$z=2.3$ & CMQ & $12.3_{-5.9}^{+7.0}$ & $9.8_{-4.9}^{+6.2}$ & $15.3_{-7.0}^{+8.0}$ & $9.1_{-4.7}^{+5.9}$ & $11.1_{-5.4}^{+6.6}$ \\
\multicolumn{7}{l}{(a) Symbols refer to: M = massive galaxies ($\log(M_*/M_\odot)>11$), C = compact galaxies ($R_e<2$~kpc), Q = Quiescent galaxies}\\
\multicolumn{7}{l}{ ($\log({\rm sSFR/yr}^{-1})<-11$) and combinations thereof.} \\
\multicolumn{7}{l}{(b) The co-moving volume for the $z\gtrsim3$ and $z=2.3$ populations is computed adopting the redshift range $2.8<z<4.0$}
\end{tabular}
\end{center}
\end{table*}

\subsection{The evolution of massive $z>3$ galaxies to $z\sim2$}

In order to explore how our results depend on different SFH in SED modelling, we adopt a different SFH for the past and future history of galaxies in each panel of Figure \ref{fig:SFHs}. Specifically, for the exponential and truncated exponential SFH, we used stellar masses and SFRs computed adopting an exponential SFR and a \citet{salpeter1955} IMF, and converting the SFRs and stellar masses to a Kroupa IMF by subtracting 0.2 dex. The location of the starting galaxies does not depend much on the SFH used to compute the stellar masses and SFRs, meaning that the choice of SFH primarily effects the evolved galaxies.

The number density $n_{M}$ of massive ($M_*>10^{11}M\odot$) $z>3$ galaxies from our sample is $n_{M}=14.9_{-6.8}^{+7.9} \times 10^{-6}$~Mpc$^{-3}$; Figure \ref{fig:numden} shows that it is fully compatible with the number density of $z\sim2.2-2.5$ quiescent galaxies from \citet{brammer2011}. In particular, this implies that the whole population of massive $z>3$ galaxies needs to have its SFR  quenched to $\log({\rm sSFR/yr}^{-1})<-11$ by $z\sim2$.

The comparison of the number density of $z>3$ massive galaxies with $z=2.3$ massive, quiescent galaxies, both computed using the full NMBS catalog, confirms the above result.

This is further supported by our SFH predictions. The blue bar in the left panel of Figure \ref{fig:numden} represents the number density of galaxies which, evolved from their observed redshifts, are massive and quiescent by $z=2.3$. The box takes into account the spread in values from 4 of the 5 adopted SFHs (excluding the CSF), and cosmic variance.  The blue box is in good agreement with measurements of quiescent galaxies from \citet{brammer2011} at $z\sim2-2.5$. Over-plotted are also the values from the individual SFHs. In general, apart from the constant SFH case, it is not possible to discriminate amongst the SFHs, mainly due to the high cosmic variance errors ($\approx 0.35$ relative error). As shown in the left panel of Figure \ref{fig:numden}, 4 out of 5 of the SFHs do equally well in reproducing the number counts of MQGs at $z\sim2.3$. The only exception is the CSF that  underpredicts the observed number density of MQGs at $z\sim2$ by a factor of $\sim4$. 
This suggests that the only valid SFHs are the declining or truncated SFHs and that the constant SFH  is not a valid option in this range of redshift. 
On the other side, given that the number densities of massive, star-forming galaxies at $z\sim2$  is approximately the same as that of MQGs \citep{brammer2011}, the fact that the CSF SFH underpredicts the number densities of MQG means that the CSF SFH would underpredict also the number density for the massive star-forming galaxies at $z\sim2$. This may suggest that high-redshift massive star-forming galaxies are characterized by rising SFHs, in agreement with recent works (e.g. \citealt{maraston2010,papovich2011}).

Interestingly, \citet{bell2012} found a similar result when comparing $z\sim1$ galaxies to the $z=0$ quiescent population, with the local population of MQGs  approximately as numerous as the entire massive population at $z\sim1$. This could suggest that mechanisms for quenching, either the same over the history of the Universe or of different nature at different epochs, are ongoing more or less continuously since $z\sim4$. 

\subsection{The progenitors of $z\sim2$ compact, massive, quiescent galaxies}

\label{sec:cmq}

Several works have demonstrated that the vast majority of massive quiescent $z\sim2$ galaxies are also compact (see \citealt{szomoru2012} and references therein). This allows us to directly compare our number densities for compact, massive, quiescent galaxies to the number densities at $z=2-2.5$ from  \citet{brammer2011}. 

\citet{bezanson2009} published an estimate of the number density of compact, massive, quiescent galaxies, based on the stellar mass function of \citet{marchesini2009}. The corresponding value is marked in Figure \ref{fig:numden} by the crossed-circle symbol. We increased the original error bar to take into account the effects of cosmic variance using the \citet{moster2011} recipe, and adding it in quadrature.  The final error bar is a factor of $\approx1.5$ the error bar quoted in the original work, but more representative of the true uncertainties.

As is seen in the right panel of Figure \ref{fig:numden}, the number density $n_{C}$ of massive galaxies that are compact at $z>3$ (which also corresponds to the number density $n_Q$ of massive quiescent galaxies at $z>3$)  is a factor of $\sim3$ smaller compared to the number density of all massive galaxies. Their evolution to $z=2.3$ using our chosen SFHs (blue box) produces a number density which is approximately a factor of $\gtrsim 4$ smaller than the number density of massive quiescent galaxies from \citet{brammer2011}, and consistent only at a 2$\sigma$ level. We can explain the decrease in number density between $z\sim3$ and $z\sim2$ in the context of our evolution models: the increase with cosmic time of the fraction of mass returned to the ISM decreases the stellar mass. One possible source of the discrepancy between our measured number density and \citet{brammer2011} could be the different criteria adopted by \citet{brammer2011} to select the quiescent population (the UVJ color selection - \citealt{williams2009}). In a recent work, \citet{szomoru2012} showed that the fraction of quiescent galaxies selected at $z\gtrsim2$ using the UVJ method is equivalent to a sSFR$<10^{-10}$yr$^{-1}$ criteria\footnote{We also selected quiescent galaxies in the full NMBS data set using the UVJ color-color technique. The sSFRs for quiescent galaxies at $2<z<2.5$ and with $\log(M_*/M_\odot)>11$  are lower than log(sSFR/yr$^{-1}$)$=-9.84$, with a 75\% upper limit  log(sSFR/yr$^{-1}$)$=-10.3$ and a median log(sSFR/yr$^{-1}$)$=-11.11$ }. It is worth noting that we do not find $z>3$ compact, massive, star-forming galaxies, and specifically compact, massive galaxies with $-11<\log($sSFR/yr$^{-1}$$)<-10$ which, following our evolutionary models, would eventually become compact, massive, quiescent by $z\sim2$. This means that our measurement of the number density of compact, massive, quiescent galaxies at $z\sim2.3$   does not change if we use a sSFR that is compatible with that for a UVJ cut.

When we compare the expected number density of CMQ galaxies that have been evolved from $z\sim3$ with the measured sample of $z=2.3$ MQ galaxies from NMBS (nearly all of which are compact), we find that the two agree to within one sigma.

 The plot shows also that the number density from \citet{bezanson2009} is higher than our measurements, and compatible only with the upper end of the \citet{brammer2011}. One possible reason for this discrepancy could be that  \citet{bezanson2009}  assume the fraction of massive, quiescent galaxies to be 0.5, likely a too optimistic choice (e.g. \citealt{dominguez-sanchez2011}).

Similar to the left panel, in the right panel of Figure \ref{fig:numden} the number densities from the individual SFHs are marked. For the compact massive galaxies, the number density from the CSFH is compatible with the measurements from the other SFHs, in agreement with recent works (see e.g. \citealt{gonzalez2010,reddy2012}).

\subsubsection{The effects of mergers}
In order to try to understand the possible mechanisms which could be responsible for the build up of the compact, massive, quiescent population at $z\sim2$, we implemented a very simple statistical model for how mergers could affect the evolution in the observed number densities. Our toy model consists of randomly choosing pairs of galaxies among the 110 galaxies with $z>3$ and with a measurement of the $R_e$ and leaving the SFH of each galaxy to evolve independently of its companion. The stellar mass of the merged pair was finally considered as the algebraic sum of the two components; the sSFR was computed as the mass-weighted sum of each component (or, equivalently, as the sum of the two SFRs divided by the total stellar mass), while we assumed for the $R_e$ of the merged pair the $R_e$ of the more massive companion.

The effect of merging on the number density evolution are represented in the right panel of Figure \ref{fig:numden} by the red box; the result is an agreement between the observed number density of compact MQGs at $z\sim2$ and that from our models, for all the SFH but the CSF. In this case, in fact, the number density is even smaller than the number density of the population without merging. In the framework of our toy model, this is due to the high probability that each pair contains a galaxy with high SFR. The end product is then biased towards high sSFR galaxies at the end of the merging process. 

Given the potential effect of a starburst or AGN that is triggered by the merger, we also examined a scenario in which the SF is quenched following the merger. This is supported by recent hydrodynamical simulations  in which mergers play an important role in the gravitational heating of the halo gas, and consequently in the suppression of star formation (\citealt{johansson2009}). However we would like to note that this last process does not improve the agreement significantly, except for the CSF case, for which the number density falls well within the values from the other SFH.

Good agreement between the expected and observed number densities for our evolved, merged, and quenched model is also found when we compare the expected number densities to those calculated from the full NMBS sample using a $\log($sSFR/yr$^{-1})<-11$ cut.  This comparison is useful as the full NMBS catalog has a higher precision than our sample that is limited to the area with CANDELS overlap.  Although the merging model appears to be in better agreement with the data, it is clear that even a simple evolution model with no merging provides an adequate match to the data.  Therefore, merging and quenching, which are both likely processes that are occurring in the galaxy population, are not required to explain the number densities.\\

If the two quiescent galaxies with MIPS detections indeed have the extreme SFRs implied by the rest-frame 5.6~$\mu$m fluxes, our results on the potential importance of merging would remain qualitatively unchanged.  This is because these two galaxies are only 13-20\% of the whole sample of galaxies that would be classified as massive at $z\sim2.3$.  If the truncated post-merger SFH is correct then the results will also remain quantitatively similar as the prompt quenching of star formation assumed in this model would turn them into massive quiescent galaxies by $z\sim2.3$.

\section{Discussion}

According to current semi-analytic models for the formation of massive quiescent galaxies,  the bulk of the stellar mass was formed as a compact, massive spheroid at $z>3$ through gas-rich merger events (e.g. \citealt{oser2010}). Specifically, in the redshift range between $z=5$ and $z=3$, the models predict that the central galaxy would still be building up from gas flows which would feed the the formation of stars in the central region of the galaxy directly, forming the concentrated stellar system \citep{naab2009}. In a second stage, quenching mechanisms such as major merger or feedback from AGN or from star formation would convert the full population of massive galaxies into the population of quiescent massive galaxies observed at later cosmic times. Subsequent minor dry mergers would be responsible for the increase in size \citep{naab2009,oser2012}, while keeping the gain in stellar mass to a factor of $\lesssim2$ \citep{vandokkum2010}. 

Understanding \emph{what} are the physical mechanisms of star-formation quenching, \emph{when} they started to act and \emph{how long} they took to completely quench star-formation is then a central key in our knowledge of formation and evolution of local massive quiescent galaxies. While our analysis does not provide us with significant information on the physical mechanisms, it allows us to put new constraints on both when star-formation quenching could have happened and on how fast it could be.

In fact, the compatibility of the number density of massive $z>3$ galaxies with the number density of $z\sim2$ massive quiescent galaxies from the literature suggests that the population of massive quiescent galaxies at $z\sim2$ can be completely accounted for by the observed $z>3$ population of massive galaxies that is subsequently quenched. Specifically, when evolving our sample of $z>3$ galaxies to $z=2.3$, a good agreement is obtained introducing quenching of star-formation after merging, process which is supported by recent hydrodynamical simulations.

Secondly, our observations of compact, massive, quiescent galaxies at $3<z<4$ push back in time the appearance of this class of objects to when the Universe was $\sim2$ Gyr old.

Finally, considering that, according to models, massive galaxies should still be actively forming stars at $3<z<5$, then the  observation of compact massive galaxies which are already quiescent by $3<z<4$ imposes that the quenching of star formation should be a rapid mechanism in massive galaxies, acting on timescales of less than 1~Gyr in the early Universe.

\section{Conclusions}

In this work we used two overlapping public sets of data in a region of the COSMOS field to identify the progenitors at $z\gtrsim3$ of the compact, massive, quiescent galaxies observed at $z\sim2$. Stellar masses, sSFRs and photometric redshifts were taken from the NMBS. The sizes were measured on high-resolution CANDELS F160W images. The population of $z\gtrsim3$ galaxies was evolved to $z=2.3$ using \cite{bruzual2003} models following five different SFHs (constant, delayed-exponential, truncated delayed-exponential, exponential and truncated exponential). All the number densities were computed assuming a co-moving volume corresponding to $2.8<z<4.0$. Our main results can be summarized by the following points:

\begin{enumerate}
\item We discovered four compact, massive ($M_*>10^{11}M_\odot$) quiescent (SED-based sSFR$<10^{-11}$~yr$^{-1}$) galaxies at $z\gtrsim3$, corresponding to a completeness-corrected number density of $n_Q=5.2_{-3.1}^{+4.6}\times10^{-6}$~Mpc$^{-3}$. If the two galaxies with MIPS detection were excluded, the resulting completeness-corrected number density would be $n_Q=3.6_{-2.4}^{+4.0}\times10^{-6}$~Mpc$^{-3}$.
\item For a complete sample of 10 galaxies with $\log M_*/M_\odot>11$, we found that the quiescent (sSFR$<10^{-11}$ yr$^{-1}$) galaxies are compact ($R_e\sim1.2$ kpc), while the star-forming galaxies are extended ($R_e \sim 3.1$ kpc), qualitatively similar to what is found at lower redshifts ($z\sim2.3$). If the two quiescent galaxies with MIPS 24~$\mu$m detection were considered star forming, the averaged sizes of the quiescent and star-forming galaxies would be $R_e=0.6$~kpc and $R_e=2.8$~kpc, respectively. 
\item We found five compact ($R_e<1.4$~kpc), star-forming (sSFR$ \sim 10^{-9}$~yr$^{-1}$) galaxies at $z\sim3$ with  large stellar masses ($10^{10.6}<\log(M_*/M_\odot)<10^{11.0}$). The small effective radius was confirmed by a \citet{sersic1968} profile fitting of the stacked image. 
\item  The number density of massive $z>3$ galaxies is comparable to the number density of $z\sim2$ massive quiescent galaxies from the literature. The evolution of the number density of the $z>3$ galaxy population to $z\sim2$ can be accounted for with a family of decaying or truncated SFHs.  The CSF SFH does not fit the observed number densities.  A
model with quenching of the SFR between $z=2.3$ and 3 does.
\item When we evolve our $z>3$ galaxies to $z=2.3$, we find
that the predicted number density of compact, massive, quiescent
galaxies is consistent at the $1-\sigma$ level for all of our adopted SFHs. An even better agreement is obtained if we quench the star formation after the merging event.  Such a rapid truncation is supported by recent simulations (e.g. \citealt{johansson2009}).
\end{enumerate}

The above results lead to the following conclusions:

\begin{itemize}
\item  The population of massive quiescent galaxies at $z\sim2$ can be completely accounted for by the observed $z>3$ population of massive galaxies that is subsequently quenched.
\item  The existence of compact, massive, quiescent galaxies at $3<z<4$ pushes back in time the appearance of this class of objects to when the Universe was $\sim2$~Gyr old. 
\item Since galaxy formation models predict that at $3<z<5$ galaxies should still be building up from gas flows, the existence of compact, massive galaxies that are quiescent already at $3<z<4$ implies that the quenching of star formation should be a rapid mechanism, acting on timescales of less than 1~Gyr in the early Universe.
\end{itemize}

We note that half of the sub-sample of massive and quiescent galaxies is detected in MIPS, implying large SFRs if the MIR emission were associated to dust-enshrouded star formation. Since the MIPS band probes rest-frame wavelengths shorter than 6 $\mu$m at $z>3$ (i.e., emission from hot dust), SFRs derived from MIPS are very uncertain and can potentially be contaminated by emission from a dusty torus of an AGN. Observations in the far-IR (e.g., ALMA) are needed to robustly quantify the level of obscured star formation and to confirm the quiescent nature of these galaxies.

The analysis presented in this work is based on a sample size of about one hundred objects, resulting from the intersection of two catalogs, CANDELS and NMBS on the single COSMOS field. Recent projects like the 3D-HST survey \citep{vandokkum2011,brammer2012} will provide accurate redshift for $\sim7000$ objects at $1<z<3.5$ over a total area a factor of $\sim5$ larger than the one available for this work. This will reduce by a large amount both the Poisson noise and the uncertainties due to the cosmic variance. Additionally, the increase in photometric depth from both CANDELS and the 3D-HST with respect to the currently available data sets will also allow us to probe the population of galaxies down do smaller stellar masses. These improvements will contribute significantly in the near future to further understanding the build-up and evolution of massive galaxies.

\acknowledgments

The authors are thankful to the anonymous referee for his/her comments and suggestions which helped improving the paper. This material is based upon work supported by the National Science Foundation under Award No. EPS-0903806 and matching support from the State of Kansas through Kansas Technology Enterprise Corporation.  
This study makes use of data from the NEWFIRM Medium-Band Survey, a multi-wavelength survey conducted with the NEWFIRM instrument at the KPNO, supported in part by NSF and NASA. This work is based on observations taken by the CANDELS Multi-Cycle Treasury Program with the NASA ESA HST, which is operated by the Association of Universities for Research in Astronomy, Inc., under NASA contract NAS5-26555. This work is based on observations made with the Spitzer Space Telescope, which is operated by the Jet Propulsion Laboratory, California Institute of Technology under a contract with NASA. This research made use of the OSX Version of SCISOFT assembled by Dr. Nor Pirzkal and F. Pierfederici. DM acknowledges the support of the Tufts University Mellon Research Fellowship.

\bibliographystyle{apj}

\end{document}